\documentclass[twocolumn, tighten, times]{aastex63}
\usepackage{amsmath,amstext}
\usepackage[T1]{fontenc}
\usepackage[figure,figure*]{hypcap}
\usepackage{multirow}
\usepackage[T1]{fontenc} 
\usepackage[utf8]{inputenc} 
\usepackage{float}
\usepackage{bigints}
\usepackage{lineno}
\usepackage{soul}
\usepackage{babel}
\usepackage{natbib}

\received{XXX}
\accepted{YYY}

\shorttitle{Collisional growth of planetesimals}
\shortauthors{Best et al.}

\defcitealias{Best2024}{\hyperlink{Best2024}{Paper I}}

\begin{document}

\title{The Influence of Cold Jupiters in the Formation of Close-in planets. II. Collisional Growth of Planetesimals}

\correspondingauthor{Marcela Best}
\email{marcela.best@usach.cl}

\author[0000-0001-8361-9463]{marcela best}

\affiliation{Universidad de Santiago de Chile, Av. Libertador Bernardo O’Higgins 3363, Estación Central, Santiago}
\affiliation{Millennium Nucleus on Young Exoplanets and their Moons (YEMS), Chile}
\affiliation{Instituto de Astrofísica, Pontificia Universidad Católica de Chile, Av. Vicuña Mackenna 4860, 782-0436 Macul, Santiago, Chile}
\affiliation{Millennium Institute of Astrophysics MAS, Nuncio Monsenor Sotero Sanz 100, Of. 104, Providencia, Santiago, Chile}
\affiliation{Department of Astronomy, Indiana University, Bloomington, IN 47405, USA}

\author[0000-0003-4623-1165]{Antranik A.  Sefilian}
\affil{Department of Astronomy and Steward Observatory, University of Arizona, 933 North Cherry Avenue, Tucson, AZ 85721, USA}

\author[0000-0002-9196-5734]{Carolina Charalambous}
\affiliation{Instituto de Astrofísica, Pontificia Universidad Católica de Chile, Av. Vicuña Mackenna 4860, 782-0436 Macul, Santiago, Chile}
\affiliation{Department of Astronomy, Indiana University, Bloomington, IN 47405, USA}

\author[0000-0003-1572-0505]{Kedron Silsbee}
\affiliation{University of Texas at El Paso, El Paso, TX 79968, USA}

\author[0000-0003-0412-9314]{Cristobal Petrovich}
\affiliation{Department of Astronomy, Indiana University, Bloomington, IN 47405, USA}
\affiliation{Instituto de Astrofísica, Pontificia Universidad Católica de Chile, Av. Vicuña Mackenna 4860, 782-0436 Macul, Santiago, Chile}

\begin{abstract}
\noindent Exoplanet observations have shown that the occurrence and orbital architectures of close-in super-Earths and sub-Neptunes are shaped by the presence of outer gas giant planets.
This influence may emerge during the formation stage or from later dynamical evolution by a yet elusive physical process.
In this work, 
we investigate the early stages of planetesimal accretion, modeling the joint collisional and dynamical evolution of planetesimals under the gravitational influence of a cold Jupiter and a viscously-dissipating massive protoplanetary disk.
We find that an initially extended planetesimal disk of small ($\sim 1-10$ km)
bodies evolves into massive, compact ($\Delta a/a\lesssim 0.1$) rings of several Earth masses in Moon-sized objects centered at $\sim0.1-0.5$ au.
This prevalent outcome is the result of an initial stage of planetesimal accretion over the first 10 kyrs, followed by orbital transport driven by a secular apsidal resonance sweeping inward on Myr timescales. 
Our findings highlight the crucial role of giant planets in redistributing solids within the inner disk.
This redistribution of planetary building blocks may help explain why systems with giant companions often depart from the “peas-in-a-pod” architecture.


\end{abstract}

\keywords{planets and satellites: formation --- protoplanetary disks --- planet-disk interactions}

\section{Introduction} \label{sec:introduction}

The effects of giant planet formation within a protoplanetary disk on the subsequent formation of planets interior to its orbit remains an open question. Radial velocity follow-up observations of transiting planetary systems suggest that systems hosting one or more giant planets are more likely to also harbor inner super-Earths or sub-Neptunes \citep{ZhuWu2018,bryan2019,correlation_2021,correlation_2025}. Moreover, there is ongoing debate on the statistical evidence for this correlation and, more recently, was thought to depend on the metallicity of the host star \citep{bonomo,vanzandt,bonomo2025}. That is, systems around metal-rich stars seem to show a stronger association between giant planets and inner low-mass companions compared to those around metal-poor stars \citep{zhu2024}. 

A related trend on the architecture of close-in planets is the ``peas-in-a-pod'' pattern,  where planets typically exhibit uniform orbital spacing \citep{original_peasinapod}. This pattern appears to disappear when the cold giant planets co-exist in the same system \citep{heweiss}. However, such a disruption may not necessarily be related to the formation of the close-in planets, as it could be the result of a post-formation dynamical upheaval \citep{zhang2021, LiveseyBecker2025}.

At the formation level,  a possible correlation between cold giant planets and the presence of inner super-Earths would contradict models of planet formation through pebble accretion \citep{pebble1,pebble2}. According to these models, the flux of pebbles migrating inward from the outer disk would be blocked by the giant planet, effectively starving the inner system of the material necessary for forming large rocky planets \citep{mulders2021} or preventing inward migration of already formed planets  \citep{izidoro2015}. However, it should be noted that some studies disagree on the effectiveness of giant planets as barriers for the flux pebbles \citep{filtering_1, filtering_2}. In particular, material accumulating at the pressure maximum exterior to the giant's orbit may grind down to smaller sizes, thus allowing it to filter inside the planet's orbit, and then reassemble interior to the planet, reconstituting the pebble flux. This effect would still decrease the pebble flux, roughly by an order of magnitude after 1 Myr compared to the case of a disk without a Jupiter-mass planets based on the calculations by \cite{filtering_2}. The actual decrease depends on many factors such as the strength of the material (e.g. rocky or icy), the viscosity at the gap, the planetary core mass creating the gap, among others. 

In this paper, we study the effects of planetary embryo formation through core accretion in the presence of an embedded giant planet, a scenario which remains largely unexplored in exoplanetary settings \citep[see][for related work in the Solar System context]{Bromley_2017}. Such an investigation was presented in our previous study \cite[][hereafter, ``{Paper I}'']{Best2024}, where we semi-analytically investigated the process of material transport within a protoplanetary disk assisted by a sweeping secular apsidal resonance. This resonance arises from a matching of the secular frequencies of the planetesimal precession rates (due to the giant planet and the gaseous disk) with that of the planet (due to the disk). As the protoplanetary disk dissipates, the resonance sweeps inward toward the star, exciting the eccentricities of planetesimals along its path to values as high as $\sim$0.1, depending on the eccentricity of the giant. The resulting increase in eccentricity enhances gas drag, which in turn causes planetesimals to migrate inward at a faster rate, leading to the transport of material and the formation of overdense rings of solids within the disk (at $\sim 0.1-0.5$ au for a gas giant at $\sim 3-5$ au). Such ring-like distributions have been proposed for the formation of terrestrial planets both in the Solar System \citep{hansen_2009,woo_2023,dale_2025} and exoplanetary settings \citep{batygin_2023}.

While \hyperlink{cite.Best2024}{Paper I} demonstrated the effectiveness of this mechanism in transporting planetesimals of various sizes (ranging from 1 to 100 km), it did not account for the collisional evolution of planetesimals. This missing physics is essential for investigating the growth of planetary cores and assessing whether the accumulation of solid material ahead of the resonance creates an optimal environment for planetary formation. 

In this work, we extend our previous study by incorporating collisional interactions among planetesimals. Our approach is based on previous related work, most notably by \cite{smoluchowski}, who developed a foundational framework to model the dynamics of particle coagulation. Naturally, this framework allows one to describe how small planetesimals collide and grow into larger bodies, a fundamental process in planet formation. Subsequent work, such as that by \cite{collisions}, expanded on these models by including the effects of fragmentation, where collisions between planetesimals may result in partial mergers, triggering a fragmentation cascade of smaller particles. This process depends strongly on the impact velocity, which is governed by the eccentricity and orbital alignment of the colliding planetesimals, as explored in \hyperlink{cite.Best2024}{Paper I}, as well as on the sizes and composition of the planetesimals. More recently, \cite{silsbee} expanded on these approaches by incorporating the orbit-averaged effects of an external stellar binary and disk gravity on the coupled dynamical-collisional evolution of planetesimals. In this work, we adopt a similar methodology but apply it to the case of an exterior giant planet rather than a stellar companion.

By incorporating collisional evolution into the model of \hyperlink{cite.Best2024}{Paper I}, our aim is to provide a more comprehensive view of how secular resonances and material transport shape planetary formation. This more self-consistent approach allows us to assess better whether the dynamical environments created by secular resonances promote efficient planet formation, while also providing insights into the processes that sculpt the architecture of emerging planetary systems.

This paper is structured as follows. In Section \ref{sec:methods}, we first provide a brief overview of the equations governing the dynamical evolution of planetesimals, and then introduce our methodology for modeling collisions among them. Section \ref{sec:results} presents our main results, where we describe the different stages in the evolution of a fiducial system and examine how various parameters, such as the orbits and masses of the giant planet, affect the results. In Section \ref{sec:discussion}, we discuss the implications of these results for the observed giant planet and super-Earth correlation as well as the resulting gap complexity. A summary of our findings and concluding remarks is presented in Section \ref{sec:conclusions}. 
Technical details of calculations are described in Appendices \ref{app:disk_evolution}--\ref{app:physical_parameters}.

\section{Methods} 
\label{sec:methods}

Our setup follows that described in \hyperlink{cite.Best2024}{Paper I} (see Figure 1 therein for a schematic overview). Namely, the model system consists of planetesimals initially on circular orbits which evolve under the combined action of a depleting gaseous disk---which contributes both gravitationally and through gas drag---and an external giant planet on a slightly eccentric orbit. The cold Jupiter's longitude of pericenter $\varpi_J$ precesses at a rate determined by the gravity of the axisymmetric gaseous disk on either side of its orbit, while its eccentricity $e_J$ and semimajor axis $a_J$ remain constant. Additionally, the entire system is taken to be coplanar. For completeness, details of the gaseous disk's viscous evolution are provided in  Appendix \ref{app:disk_evolution}.

Within this model system, the long-term secular evolution of the planetesimals' semimajor axes $a_p$, eccentricities $e_p$, and longitudes of periapsis $\varpi_p$ are governed by the following equations (\hyperlink{cite.Best2024}{Paper I}):
\begin{eqnarray}
\frac{de_{p}}{dt} &=& \left( \frac{de_{p}}{dt} \right)_{\rm drag} + \left( \frac{de_{p}}{dt} \right)_{\rm CJ} , \label{eq:1}
\\
\frac{d\Delta \varpi}{dt} &=& \left( \frac{d\varpi_{p}}{dt} \right)_{\rm disk} + \left( \frac{d\varpi_{p}}{dt} \right)_{\rm CJ} - \left( \frac{d\varpi_{J}}{dt} \right)_{\rm disk}  , \label{eq:2}
\\
\frac{da_{p}}{dt} &=& \left( \frac{da_{p}}{dt} \right)_{\rm drag} , \label{eq:3} 
\end{eqnarray}
where $\Delta \varpi \equiv \varpi_{p} - \varpi_{J}$.
In Equations (\ref{eq:1})--(\ref{eq:3}), the subscripts ``CJ'',  ``drag'', and ``disk'' refer to the contributions arising from the cold Jupiter, gas drag, and disk gravity, respectively. The individual expressions for each of the terms in these equations are provided in \hyperlink{cite.Best2024}{Paper I}; see equations (11)--(21) therein.\footnote{Note that there is an unfortunate typo in \hyperlink{cite.Best2024}{Paper I}: in equation (12), $\rho$ refers to the volumetric density of the gaseous disk, not that of the planetesimal as incorrectly stated therein.}  All terms, except those related to drag, are expanded to second order in eccentricity; for the drag-induced components, higher-order expressions are retained to capture eccentricity-dependent effects. Here, we note that a secular apsidal resonance occurs at the location where $d\Delta \varpi/dt = 0 $, which in turn forces the planetesimal eccentricities to grow to relatively large values, $e_p(t) \rightarrow 1$. As the gaseous disk dissipates, the resonance location progressively shifts inward.

As briefly discussed in Section \ref{sec:introduction}, accounting for planetesimal collisions is a necessary step to better understand the fate of the material accumulated by the sweeping secular resonance. In order to analyze such effects, we now extend the framework of \hyperlink{cite.Best2024}{Paper I} by incorporating a new module that accounts for planetesimal collisions. This allows the planetesimal population to evolve  both their number and size distributions through processes of  growth and fragmentation. 
The following subsections detail the collision module's implementation, which is largely based on the work of \citet{silsbee}. 
Readers interested in the impact on the resultant planetesimal dynamics may proceed directly to Section \ref{sec:results}.

\subsection{Collisional evolution of planetesimals}

To follow the collisional evolution of planetesimals in our simulations, we divide them into distinct mass bins. The number $N_i$ of bodies in the mass bin $i$ is updated at each time step following equation (A2) in \cite{silsbee}:
\begin{equation}
    \Delta N_i = \sum_j \sum_k C_{ijk}  ,
\end{equation}
where
\begin{equation} \label{eq:c_ijk}
    C_{ijk} = {\rm Poisson}(\mathcal{R}_{jk}\Delta t) \left(\frac{1+\delta_{jk}}{2}\right) \left(F_{ijk}-\delta_{ij} - \delta_{ik}\right)   .
\end{equation}
Here, particles $j$ and $k$ collide with each other to produce particle $i$, where $\mathcal{R}_{jk}$ is the collisional rate (Eq. \ref{eq:rates}), $F_{ijk}$ describes the size distribution of the collisional outcomes  (Eq. \ref{eq:cascade}), $\Delta t$ is the time step (Eqs. \ref{eq:local} and \ref{eq:global}), and $\delta$ is the Kronecker delta function. Each of these terms is described in more detail in the following.

\subsubsection{Collisional outcomes: generic}
\label{sec:col_outcome_s}

Generally speaking, there are two extreme outcomes when two planetesimals of any given size collide. One outcome is that the  planetesimals could stick together,  forming a single, larger planetesimal with a total mass equal to the sum of the two colliding masses ($m_{\rm tot} = m_1 + m_2$). A second outcome would be the complete fragmentation of the colliding planetesimals. That is, the collisional outcome would be a cloud of dust and pebbles. Since collisional outcomes form a spectrum, the actual result of a collision depends on many factors---such as encounter velocity, planetesimal size, and composition---and generally falls somewhere between these two extreme cases.

One of the main factors that determines the specific outcome of a collision is the relative velocity, ${ v_{\rm rel}}$, between the two colliding planetesimals. We quantify this as:
\begin{align}
\label{eq:v_rel}
    { v_{\rm rel}} = \lambda e_{\rm rel} {v_K},
\end{align} 
where $v_K$ is the Keplerian velocity, $\lambda$ is a number between 0.5 and 1 taken from a distribution of encounter velocities \citep[][Eq. A12 therein]{silsbee}, and the relative eccentricity is defined as:
\begin{align}
    e_{\rm rel} = | \vec{e}_j - \vec{e}_k + \vec{e}_{\rm rand} |,
    \label{eq:vec_e_rel}
\end{align} 
where $\vec{e}_{j}$ is the eccentricity vector $(e_{j}\cos\varpi_{j},e_{j}\sin\varpi_{j})$ of the colliding planetesimal $j$ (the same for $k$). The eccentricity vector corresponds to the solution of Eqs. (\ref{eq:1})--(\ref{eq:3}), which, at a given time $t$, is taken to be the forced eccentricity vector, motivated by the fact that planetesimals rapidly settle into these values (see \hyperlink{cite.Best2024}{Paper I}). 
In Equation (\ref{eq:vec_e_rel}),
$\vec{e}_{\rm rand}$ represents the dispersion around the forced value, with its norm drawn from a Rayleigh distribution with dispersion $\sigma_e$ and its apsidal orientation drawn from a uniform distribution between [0, 2$\pi$]. Thus, in cases with near-zero forced eccentricity (e.g., a circular planet or a small planet-to-disk mass ratio), $e_{\rm rel} \approx |\vec{e}_{\rm rand}|$.

Having defined the relative velocity, the associated collisional energy per unit mass can be obtained via:
\begin{align}
    Q_{R} = \frac{1}{2} \mu v_{\rm rel}^2,
\end{align} 
where $\mu = m_1 m_2 / (m_1 + m_2)$ is the reduced mass of the two colliding planetesimals. Following the collision, part of the combined mass is retained by the largest remnant, with mass $m_{\rm lr}$, while the remaining mass, $m_{\rm tot} - m_{\rm lr}$, is dispersed into a cascade of fragments.
To estimate the relative proportion of mass distributed between them, we use the following relation \citep{collisions}:
\begin{equation}
    \frac{m_{\rm lr}}{m_{\rm tot}} = 1 - \frac{1}{2} \left( \frac{Q_{R}}{Q_{\rm *}}\right), 
\end{equation} 
Here, $Q_{\rm *}$ is the threshold value above which collisions lead to fragmentation, specifically, when half of the total mass is dispersed in a fragmentation cascade. This threshold is given by:
\begin{equation} \label{eq:Q*}
    Q_{\rm *} = \left[ q_s R^{9\widetilde{\mu}/(3-2\phi)} + q_g  R^{3\widetilde{\mu}} \right] v_{\rm rel}^{2 - 3\widetilde{\mu}},
\end{equation}
where $R$ is the radius of a spherical planetesimal with the combined mass of both colliding objects. Following equation (2) in \cite{collisions}, the numerical factors for "strong rocks", such as basalt, are as follows: $q_s$ = 7$\times 10^4$, $q_g$ = $10^{-4}$, $\widetilde{\mu}$ = 0.5, and $\phi$ = 8 (in cgs units).

Once the largest remnant's mass is obtained, the remaining mass $m_{\rm tot} - m_{\rm lr}$ is distributed as a cascade of fragments. This cascade can be described as a power-law function of mass \footnote{\cite{cascade_2} find that the exact slope of this function does not significantly affect the steady state equilibrium of the mass distribution.}\citep[see Equation (6) in][and references therein]{cascade}:
\begin{equation} \label{eq:cascade}
    F(m) \propto
    \begin{cases}
        m^{-1}, &\text{\rm for }  m_{\rm min} \leq m \leq m_{\rm cutoff} \\
        0,           & \text {\rm otherwise}
    \end{cases}
\end{equation} 
with an upper mass limit of
\begin{align}
    m_{\rm cutoff} =  \begin{cases}
        0.5 (m_{\rm tot} - m_{\rm lr}), &\text{\rm for }  m_{\rm lr} / m_{\rm tot}  \geq 0.5 \\
        {\rm max}(b\cdot m_{\rm tot}, 0.5 m_{\rm lr}),           & \text {\rm otherwise}
    \end{cases}
\end{align} 
and a lower mass limit of $m_{\rm min}$, with $b$ an adjustable parameter controlling the case for destructive collisions. In our calculations, following \cite{silsbee}, we adopted $b = 0.01$. We consider a destructive collision when $m_{\rm lr} < m_{\rm cutoff}$. When this occurs, the entire available mass ($m_{\rm tot}$) is allocated to the cascade, so $m_{\rm lr}$ is set to zero. The terms $m_{\rm lr}$ and $F$ together make up the term $F_{ijk}$ in Eq. (\ref{eq:c_ijk}) when considering each pair of possible collisions.

Due to resolution constraints and limited computational resources, the cascade has a lower mass threshold ($m_{\rm min}$), below which planetesimal are no longer tracked.  In our fiducial calculations (see Table \ref{table:models}), $m_{\rm min}$ corresponds to planetesimals smaller than $1$ km. The mass lost as a result of this scales linearly with the total mass of the cascade,  $m_{\rm tot} - m_{\rm lr}$ \citep[see Eq. A11 in][]{silsbee}, and becomes particularly significant in scenarios where most collisions are destructive. We test the sensitivity of our results to this cutoff in Appendix \ref{app:resolution}, finding insignificant changes to the amount of solid material converted to large planetesimals, except in the case where $m_{\rm min}$ is comparable to the corresponding mass for the initial size of planetesimals; see panel C in Figure \ref{fig:parameters_resolution}. 

\subsubsection{Mass--semimajor axis grid and collisional interactions}

In our calculations, the disk is divided radially into $N_r$ logarithmically spaced annuli (or rings), each containing $N_m$ mass bins also distributed logarithmically. The fiducial values of these parameters, along with others, are provided in Table \ref{table:models}. 

Collisional interactions are considered in two ways. First, all mass bins within a given ring are allowed to collide with one another, following the method described thus far. Second, because some planetesimals may develop relatively high eccentricities (especially at the secular resonance peak, where $e_p\gtrsim 0.1$; \hyperlink{cite.Best2024}{Paper I}), we also allow for collisional interactions among mass bins representing different rings (i.e., at different semimajor axes). Since the number of potential interactions scales as $N_m \times N_r$, we first check whether orbits intersect before computing the collision rates (discussed next in Section \ref{subsec:col_rates_sec}).\footnote{We remind the reader that planetesimals of different sizes within a given ring have different eccentricities and apsidal angles due to gas drag, especially close to the secular resonance; see \hyperlink{cite.Best2024}{Paper I}.}

Orbital intersections are determined using equations (4)--(6) in \cite{whitmire}. These equations define the condition under which two orbits, with orbital elements labeled as `1' and `2', intersect as follows:
\begin{equation}
  (e_1 p_2)^2 + (e_2 p_1)^2 - 2 e_1 p_2 e_2 p_1 \cos(\varpi_1 - \varpi_2) \geq (p_2 - p_1)^2 ,
\end{equation}
where $p_i = a_i(1-e_i^2)$ is the semi-latus rectum  for each involved planetesimal orbit (with $i=1, 2$). The relative velocity at the two intersection points, normalized to the Keplerian velocity at 1 au around the central star ($M_*=M_{\odot}$), is given by:
\begin{eqnarray}
    \left( \frac{v_{{\rm rings},j}}{30 {\rm km/s}} \right)^2 &=&  \left[ \frac{e_1 \sin(\theta_j - \varpi_1 + \varpi_2)  }{\sqrt{p_1/{\rm au}}} - \frac{e_2 \sin(\theta_j)}{\sqrt{p_2/{\rm au}}}\right]^2 \nonumber \\
    &+& \left[ \frac{\sqrt{p_1/{\rm au}}}{r_j/{\rm au}} - \frac{\sqrt{p_2/{\rm au}}}{r_j/{\rm au}}\right]^2,
\end{eqnarray}
where $(r_j,\theta_j)$ denotes the position of the intersection point $j$ (not to be confused with the planetesimal index $i$), and $r_j = p_2 / (1+e_2 \cos \theta_j)$. Because this relative velocity is systematic whereas the velocity dispersion within a ring is random (see Eq. \ref{eq:v_rel}), we replace $v_{\rm rel}$ in Eq. (\ref{eq:v_rel}) by $v_{{\rm rings},j}$ when considering collisions between rings with different semi-major axes. 

\subsubsection{Collision rates} \label{subsec:col_rates_sec}

Once the outcome of a given collision is determined (as outlined in Section \ref{sec:col_outcome_s}), we estimate how often such collisions occur over time. We do this by calculating the collisional rates $\mathcal{R}_{jk}$ for the different populations of planetesimals (grouped by size and semimajor axis) using Equation (A.26) in \citet{silsbee}: 
\begin{equation} \label{eq:rates}
    \mathcal{R}_{jk} = f \mathcal{R}_{1,jk} + (1-f) \mathcal{R}_{2,jk}.
\end{equation} 
Here, $f = 1/ (1 + [e_{\rm rel} / \sqrt{\pi} \sigma'_e]^2 )$, $\sigma'_e = \sigma_e + c_1 e_H/2$ with  $e_H = [(m_j + m_k)/3M_{\star})]^{1/3}$ (the Hill eccentricity), and $\mathcal{R}_1$ and $\mathcal{R}_2$ correspond to the limits in which the encounter velocities are dominated by random and secular velocities, respectively, given by \citep{silsbee}:
\begin{align} \label{eq:R1}
    \mathcal{R}_{1,jk} = \frac{N_j N_k A_g v_K}{8\pi^{4} a_p^2 \Delta a_p} \left[ \mathcal{F} + \mathcal{G} \times \left( \frac{v_{\rm esc}/v_K}{2\sigma_e + c_1 e_H} \right)^2 \right] \nonumber \\ 
    \times \left( 1 + c_2 \frac{e_H}{2\sigma_e}\right),
\end{align}
\begin{align} \label{eq:R2}
    \mathcal{R}_{2,jk} = \frac{{\rm E}(\sqrt{3}/2) N_j N_k A_g v_K e_{\rm rel} }{4\pi^{5/2} \sigma_i a_p^2 \Delta a_p} \left[ 1 + \left( \frac{v_{\rm esc}}{\lambda v_K e_{\rm rel}} \right)^2 \right],
\end{align} 
where $N_j$ is the number of particles of species $j$, $\Delta a_p$ the width of the annuli, $\sigma_i$ the inclination dispersion ($10^{-4}$ for the fiducial case), $A_g = \pi (R_j + R_k)^2$  the cross section, $v_{\rm esc} = \sqrt{2G(m_j + m_k)/(R_j + R_k)}$, and $E(\sqrt{3}/2) \approx 1.21$. In Equation (\ref{eq:R1}), the numerical factors are $\mathcal{F} \approx 17.3$ and $\mathcal{G} \approx 38.2$ \citep{greenzweig}, and $c_1 = 4.75$ and $c_2 = 22.6$ \citep{silsbee}.

When the colliding planetesimals belong to the same mass bin, we apply the correction method to obtain the relative velocities as described in \citet[][Appendix A4]{silsbee}, which corresponds to a mass-weighted average sampling the masses uniformly within the bin. This accounts for the fact that each mass bin represents a distribution of masses, not just a single value.

\subsubsection{Time step determination} \label{sec:time_step}

In \hyperlink{cite.Best2024}{Paper I}, each planetesimal was evolved independently over the full simulation duration ($t_{\rm max} = {\rm 10  \, Myrs}$) since their evolution was dictated only by dynamical effects. With the inclusion of collisions in the current study, we adopt a sufficiently small time step to ensure that changes in the mass distribution remain gradual between successive steps. To achieve this, we implement the procedure described by \citet{silsbee},  summarized briefly below. 

First, the time step $\Delta t$ is chosen so that the change in the number of planetesimals $\Delta N_i$ of a given mass $m_i$ does not exceed a certain value;
\begin{align} \label{eq:local}
    \frac{{\Delta N_i}}{N_i} \leq \epsilon_{\rm local},
\end{align}
during the duration of a timestep.
However, this condition can become too restrictive for sizes that do not contribute much to the overall distribution in terms of mass, potentially leading to unnecessarily short timescales. Thus, a second condition is introduced,
\begin{align} \label{eq:global}
    \frac{{m_i \Delta N_i}}{\sum_i m_i N_i} \leq \epsilon_{\rm global},
\end{align} 
where the denominator is the total mass in a given annulus. We select the condition that results in the longest time step from Eqs. (\ref{eq:local}) and (\ref{eq:global}). This time step is calculated individually for each annulus and then the smallest value is applied uniformly to collisionally evolve all annuli. After this step, the semi-major axis of the planetesimals is evolved according to their migration rate due to gas drag, by the same duration as the collisional time step. Once this is done, the orbital elements of each planetesimal are evolved according to Equations (\ref{eq:1})--(\ref{eq:3}) for the same duration as the collisionally determined timestep.

In our fiducial calculations, we adopt $\epsilon_{\rm local} = 0.05$ and $\epsilon_{\rm global} = 10^{-3}$. We have explored other values within the ranges $\epsilon_{\rm local} \in [0.005$ -- $0.1]$ and $\epsilon_{\rm global} \in [10^{-5}$ -- $10^{-1}]$, finding that the results are converged for the fiducial choices (see panels E and F of Figure \ref{fig:parameters_resolution} in Appendix \ref{app:resolution}).

\subsection{Code and tests}

All of the procedures mentioned above for modeling planetesimal collisions have been implemented into a new module of our code, which has been made publicly available\footnote{\url{https://github.com/marcybest/collisions}}. This module is adaptable to different orbital elements of planetesimals and gaseous surface density profiles, both of which can evolve over time. It also supports modeling different solid materials and distinct disk regions. In addition, the gas drag and ring-ring interaction modules can be optionally disabled. To ensure its accuracy and reliability, the collision module has been validated against analytical solutions and previously published numerical results. 
An extensive report on these and other tests can be found in Appendix \ref{app:tests}, which, in first reading, may be skipped.

\subsection{Fiducial system parameters} \label{sec:fiducial}

\begin{table}[t!]
\centering
\caption{Simulation parameters
\label{table:models}}
\begin{tabular}{lll}
\tableline
\tableline

Parameter & Fiducial & Other Values \\

\tableline

Giant's mass, $M_{J} [M_{\rm Jup}]$ & 3 & [0.1 -- 5] \\

Giant's semimajor axis, $a_{J} {\rm [au]}$ & 3 & --- \\

Giant's eccentricity, $e_{J}$ & 0.05 & [$10^{-4}$ -- 0.25] \\

Inner disk viscosity, $\alpha_{\rm in}$ & $10^{-3}$ & --- \\

Outer disk viscosity, $\alpha_{\rm out}$ & $10^{-4}$ & ---\\

Initial gas mass [$M_{\Sun}$] & 0.01 & 0.05 \\

Gas-to-dust ratio & 0.05 & [0.01 -- 0.5] \\
Eccentricity dispersion, $\sigma_e$ & $2\times10^{-4}$ & $10^{-3}$ \\
Inclination dispersion, $\sigma_i$ & $10^{-4}$ & $5\times10^{-4}$ \\
Initial planetesimal size [km] & 5 & [2.5 -- 25] \\
Number of rings, $N_r$ & 25 & [15 -- 100] \\
Number of mass bins, $N_m$ & 50 & [25 -- 100] \\
Min. planetesimal size [km] & 1 & [0.01 -- 3]\\
Max. planetesimal size [km] & 1000 & [100 -- 2000]\\
\tableline
\end{tabular}
\end{table}

Unless otherwise stated, we adopt a fiducial system similar to that used in \hyperlink{cite.Best2024}{Paper I}, with parameters summarized in Table~\ref{table:models}. This setup features 
a dissipating gaseous disk (details in Appendix \ref{app:disk_evolution}) with an initial gas mass equal to 1\% of the stellar mass, taken to be $1 M_{\Sun}$. The disk viscosity is characterized by two $\alpha$ values, $10^{-3}$ inside and $10^{-4}$ outside the orbit of a $3$ Jupiter-mass giant planet placed at $3$ au
with an eccentricity of 0.05.
Further details and justification for this setup are provided
in \hyperlink{cite.Best2024}{Paper I}.

We also initialize a population of $5$ km planetesimals distributed across $N_r = 25$ logarithmically spaced rings from 0.1 au to 1.5 au.
In each ring, the solid mass is proportional to the local gaseous mass, resulting in an initial power-law density distribution $\Sigma_{\rm solids} \propto r^{-1}$. We adopted a solid-to-gas ratio of $5\%$, corresponding to a total solid mass of $\approx 9 M_{\oplus}$ inside the planet. 
The initial dispersions in eccentricity (around the forced value) and inclination (around $0^{\circ}$) are $\sigma_e=2\times 10^{-4}$ and $\sigma_i=\sigma_e/2$, respectively.

Collisions among planetesimals
produce new bodies, following the methodology described in this section. 
The size of the initial planetesimals affects both the time scale for forming large planetesimals ($> 1000$ km) and the fraction of material converted into such large bodies (see, e.g., Figure \ref{fig:parameters_giant}-B). 
Once planetesimals exceed $1000$ km in
radius, 
we remove them from the simulation and store them separately.
This avoids the severe time-step constraints such large bodies impose (Eqs. \ref{eq:local} and \ref{eq:global}), which would otherwise be 
significantly slowing down the computation.
Regardless, our model does not account for physical processes such as gravitational scattering, which become important at these sizes and affect collision rates (see Section 
\ref{sec:scattering}).

For reference, additional values explored for key parameters are listed in the last column of Table~\ref{table:models}.

\section{Results} 
\label{sec:results}

Here, we present the main results for the fiducial system (Sections \ref{sec:evol_phase_fid} and \ref{sec:mat_trans_fid}), and then explore the effects of varying the parameters of the system (Section \ref{sec:varying_params_result_S}).

\begin{figure*}[]\center \label{fig:fig1}
\includegraphics[width=18cm]{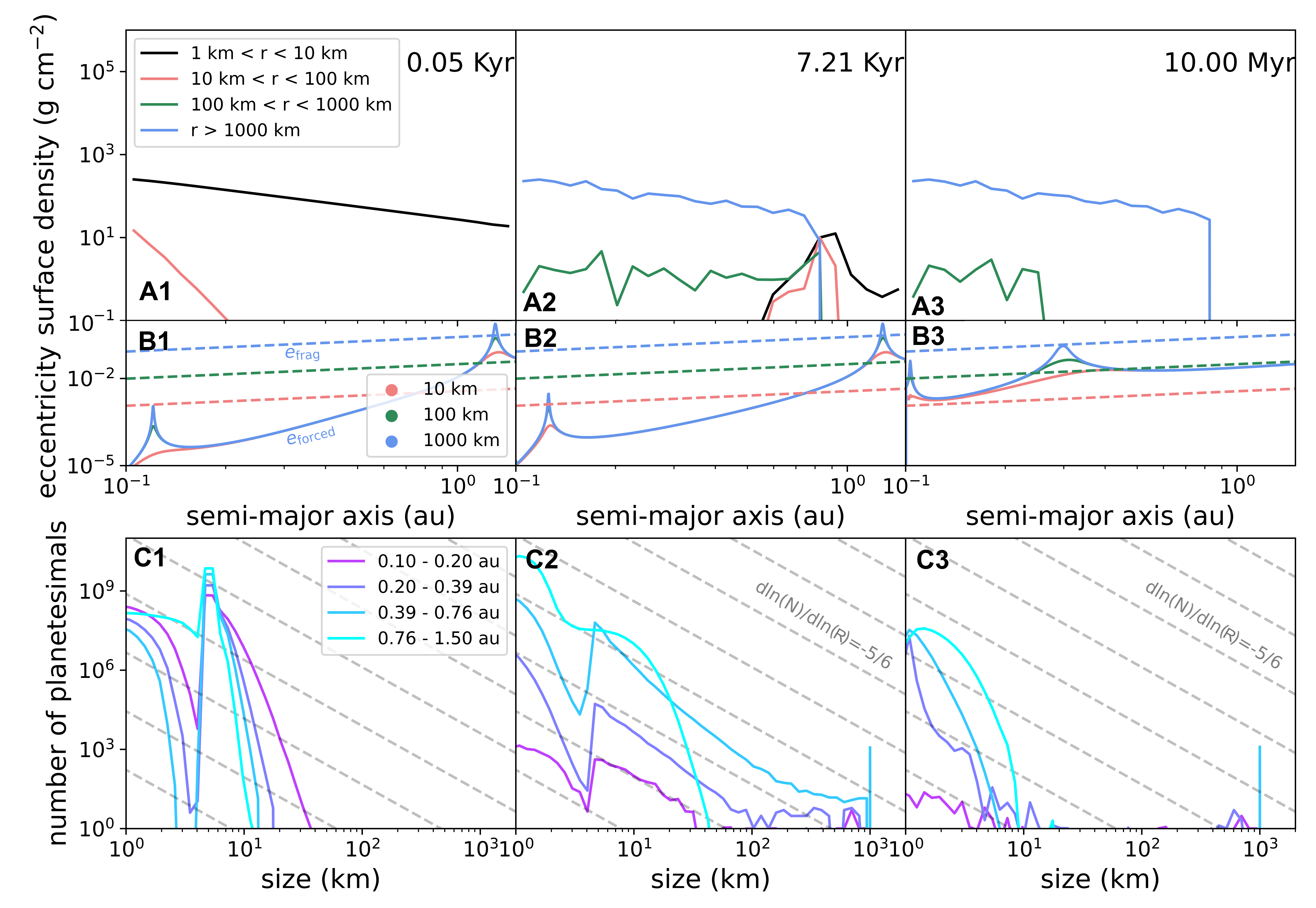}
\caption{
Evolution of the distributions of planetesimals in size and semi-major axes using the fiducial parameters (Section~\ref{sec:fiducial}) for three representative times. \textbf{Row A} presents the surface density as a function of semimajor axis for different planetesimal size ranges, indicated by different colors
as shown in the legend. \textbf{Row B} shows the forced eccentricity (solid lines) and the minimum eccentricity for fragmentation (dashed lines) as functions of semimajor axis. Different colors correspond to different planetesimal sizes. 
\textbf{Row C} depicts the size spectrum of planetesimals across four logarithmically spaced semimajor axis rings. Gray dashed lines correspond to the $-5/6$ slope of the equilibrium distribution (Appendix \ref{app:equilibrium}). Each column in the plot corresponds to a different time in the simulation, representing key stages in the evolutionary sequence as discussed in Section \ref{sec:evol_phase_fid}. An animated version of this figure is available in Figshare: \href{https://figshare.com/articles/dataset/Fiducial_Simulation_mp4/29602208?file=56386133}{10.6084/m9.figshare.29602208}. The animation runs from t = 0 to t = 10 Myrs with a duration of 60 s.}
\end{figure*}

\subsection{Evolution phases}
\label{sec:evol_phase_fid}

We ran a simulation with the fiducial system parameters (Section~\ref{sec:fiducial}) for 10 Myrs. The results are shown in Figure~\ref{fig:fig1}, with each column representing a different snapshot time. In row A, we present the surface density for four size ranges of planetesimals as a function of semimajor axis. Row B shows the radial profiles of the forced eccentricities of planetesimals alongside their fragmentation eccentricity (Eq. \ref{eq:e_frag}); the latter serves as an approximate threshold below which growth of planetesimals to larger sizes is possible.
Finally, row C shows the mass distribution across rings in different disk regions. 

Figure~\ref{fig:fig1} shows that most planetesimals larger than 1000 km form within $\sim$ 10 kyrs, as indicated by the similarity between the blue curves in panels A2 and A3.
Notably, this occurs before any noticeable sweeping of the secular resonance---as the gas mass remains close to its initial value---and before significant planetesimal migration due to gas drag, which operates on longer timescales.
Therefore, the formation of large planetesimals and the secular resonance-driven migration take place on distinctly separate timescales. Based on this, we identify three distinct phases in the evolution of planetesimals. These phases are defined by two key timescales: one associated with collisional processes and the other with the evolution of the gaseous disk (discussed in more detail below).

\begin{figure}[]\center \label{fig:fig2}
\includegraphics[width=9cm]{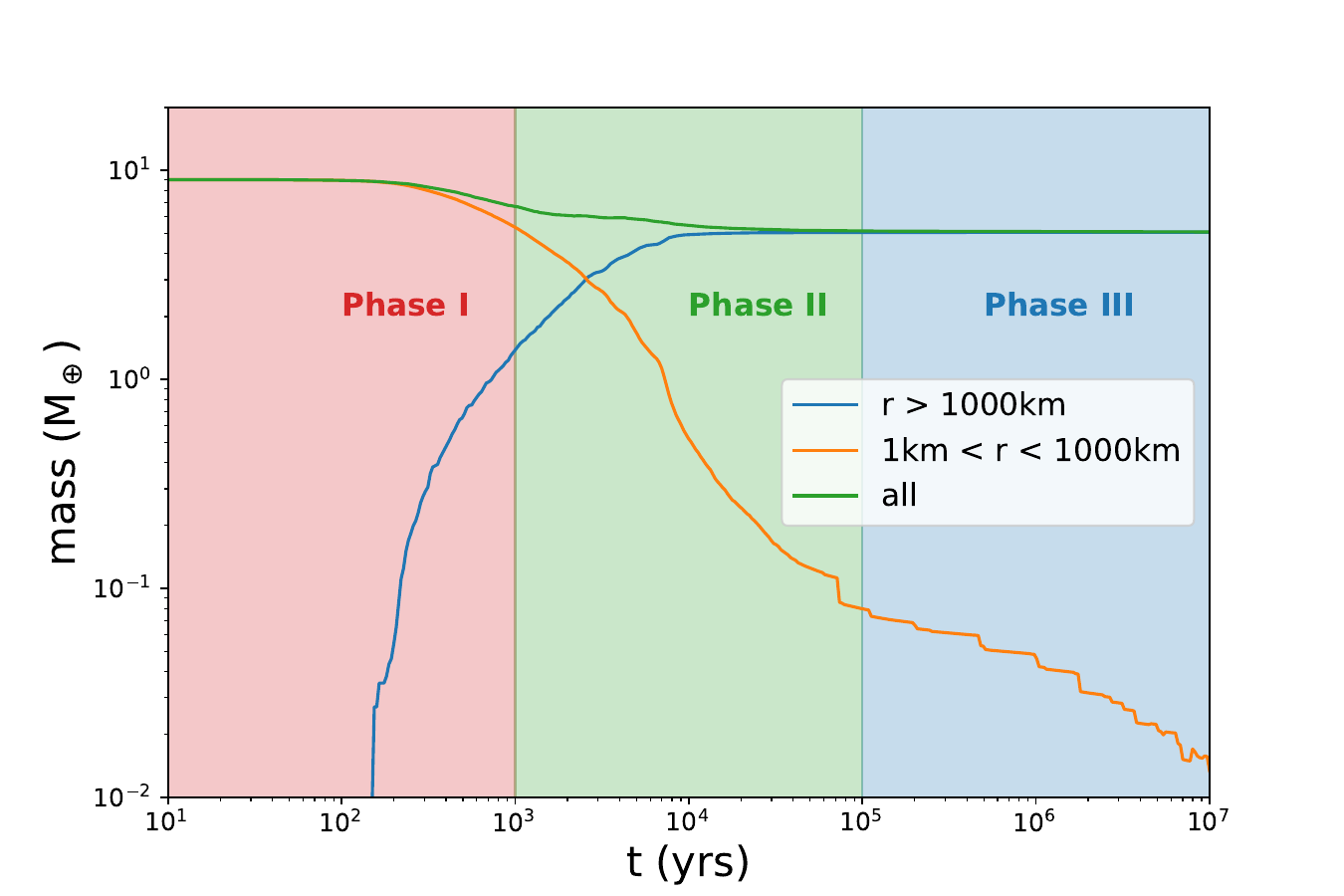}
\caption{
Planetesimal mass budget as a function of time in the fiducial simulation (Section~\ref{sec:fiducial}).  The total mass of planetesimals larger than 1000 km is shown in blue, 
those between 1 and 1000 km in orange, and their combined total mass is represented by the green curve. 
Shaded regions highlight the distinct evolutionary phases described in Section \ref{sec:evol_phase_fid}.}
\end{figure}

In Figure \ref{fig:fig2}, we identify three different evolutionary phases (indicated by shaded regions) by examining the mass evolution of planetesimals greater than 1000 km and those between 1 and 1000 km.
Phase I (in red) corresponds to the period during which nearly all large planetesimals (>1000 km) form, within the first $\sim 10^2 - 10^4$ yrs. During this time, the mass distribution transitions from a delta function to an equilibrium power-law, as shown in row C of Figure \ref{fig:fig1}.
Phase III (in blue) corresponds to the viscous timescale, by which time the secular resonance has undergone significant sweeping. At this stage, 
the large majority of the total remaining planetesimal mass resides in large ($> 1000$ km-sized) planetesimals
(see the blue and orange curves).
Between these two stages, we define an intermediate phase II (in green), during which most smaller bodies are lost to fragmentation. 
This also explains why the total mass of planetesimals (shown by the green curve) decreases most significantly during this stage: collisions grind down many planetesimals, some to below $1$ km in size, at which point they are removed from the simulation (Section~\ref{sec:methods}). 

In what follows, we discuss and quantify each of these three phases in more detail.

\textbf{Phase I - Early formation: $0 \lesssim t \lesssim t_{\rm acc}(a_p = a_{\rm trunc})$}

The first stage begins with a monodisperse planetesimal size distribution. These initial planetesimals then collide with one another, forming smaller planetesimals by fragmentation, and larger ones via mergers.
During this phase of evolution, collisions are abundant because of the large amount of available material. 
Because the collision rates are generally higher closer to the inner edge of the disk and the eccentricities are low enough, these rings form bigger planetesimals early on as can be seen from panel A1 in Figure \ref{fig:fig1}.
Relatedly, the mass spectrum deviates away from the initial delta function (more so closer to the disk's inner edge) such that it is steeper than, but has not yet attained,
the equilibrium slope of -5/6 (Appendix \ref{app:equilibrium}).

The duration of this phase can be estimated using the doubling mass timescale for a planetesimal to become 1000 km in size. Assuming a collision rate of $\Gamma = \mathcal{R}_{2,jj}$ from Equation (\ref{eq:R2}) with j corresponding to a size of approximately 800 km and without considering the gravitational focusing term, this yields:
\begin{eqnarray} \label{eq:timescale_formation}
    &&t_{\rm acc} \sim t_{\rm doubling} \approx \frac{1}{\Gamma} \sim 10 {\rm ~kyr} \left( \frac{a_p}{1 \rm au} \right)^{3/2} \left(\frac{\sigma_i}{10^{-4}} \right)  \nonumber \\
   && \times \left( \frac{\rho_s}{3\rm g/cm^3} \right)^{2} \left( \frac{50 \rm g/cm^2}{\Sigma_{\rm solids}} \right)^2 \left( \frac{2\times10^{-4}}{e_{\rm rel}} \right) \nonumber \\ 
   && \times \left( \frac{R_{\rm f}}{800 \rm km} \right)^{4} \left( \frac{M_*}{M_{\odot}} \right)^{-1/2},
\end{eqnarray}
assuming that planetesimals are spherical bodies with material density $\rho_s$, and that the relative eccentricity is dominated by $\sigma_e$ over the forced eccentricity (see left panel of Figure \ref{fig:Qstar}) so we take $e_{\rm rel} = \sigma_e$. Also, as most of the mass is concentrated in larger planetesimals (considering the equilibrium distribution for planetesimals $N\sim R^{-5/6}$), we estimate the number of planetesimals as the total solid mass contained in the annuli divided by the mass of the individual planetesimal, which slightly underestimates the doubling timescale. The numerical estimate given by Equation (\ref{eq:timescale_formation})--- which assumes equal-mass mergers--- is in good agreement with the timescales we observe in the simulations (see Figure \ref{fig:fig2}).

We note that the accretion timescale increases with semi-major axis, implying that the formation of 1000 km planetesimals proceeds from the inside out, creating a wave of growth that advances outward until it reaches some ``truncation'' radius (see outer edge of blue in 2 of Figure \ref{fig:fig1}) 
This truncation radius is mainly dictated by the position of the secular resonance. As eccentricities increase to large values near the resonance, there will be a point in semimajor axis where the forced eccentricities exceed the maximum eccentricity that leads to accretion, rendering collisions destructive.  
Over time, the truncation radius shifts inward as the gaseous disk depletes. However, we find that its initial position is the key factor in determining how far out
large planetesimals can grow within the first $\sim$1000 years. On this timescale, which is much shorter than the viscous timescale, the truncation radius remains effectively fixed.

We define the truncation radius $a_{\rm trunc}$ as the location beyond which the formation of large planetesimals halts. The exact location where this happens is set by the initial location of the secular resonance and its width. This will 
vary depending on the giant planet's properties and the 
disk's initial gas distribution (which determines the size-dependent forced eccentricities, \hyperlink{cite.Best2024}{Paper I}), but not on viscosity,  
as the disk does not have time to dissipate on these timescales. We estimate $a_{\rm trunc}$ 
as the solution to the condition given by
\citep{kobayashi2001, costa2024,sefilian_stirring}: 
\begin{eqnarray}\label{eq:e_frag}
    e_{\rm forced}^{t=0}(a_{{\rm trunc}}) &\geq& e_{\rm frag}({a_{\rm trunc}})  = 
    \frac{v_{\rm frag}(R)}{1.4 v_K(a_{\rm trunc})} , \\ \nonumber
    &\approx& 4.2 \times 10^{-4} \left( \frac{a_{\rm trunc}}{1 \rm au}\right)^{1/2} \\ \nonumber
    &\times&\left( \frac{R}{1 \rm km}\right)^{0.93} \left( \frac{M_*}{1 M_{\odot}}\right)^{-1/2},
\end{eqnarray}
which is obtained using the collisional properties of gravity-dominated bodies composed of basalt. 
This provides a good estimate for the position of the truncation radius seen in our simulations. In Figure~\ref{fig:fig1}, this truncation radius can be seen for our fiducial case (panels A2 and A3) and corresponds to the semi-major axis where the blue line (corresponding to planetesimals larger than 1000 km) drops to zero. In Row B of the same figure, the truncation radius roughly corresponds to the intersection between the solid lines ($e_{\rm forced}$) and the dashed lines ($e_{\rm frag}$) for 10 km planetesimals, which are close in size to the initially $5$ km-sized planetesimals.

\textbf{Phase II - Transient equilibrium: $ t_{\rm acc} (a_p = a_{\rm out}) \lesssim t \lesssim t_{\rm viscous}$}

At this stage, planetesimals larger than 1000 km have already formed, and the mass spectrum within each ring evolves towards an equilibrium distribution, characterized by a power-law with index -5/6 (Appendix \ref{app:equilibrium}). However, this is not the case for rings near the secular resonance, where the slope is expected to be steeper, according to Eq.~(\ref{eq:slope_v}). We find that not all rings reach this equilibrium before running out of mass, as planetesimals grow beyond the imposed upper and lower mass limits. 

As planetesimal mass continues to deplete, two distinct size populations of planetesimals emerge: the largest bodies that have been removed from the simulation and the smaller ones in the inner regions that persist but in very low numbers. 

This process occurs until $t \sim t_{\rm viscous}$, which characterizes the time it takes for the resonance location to sweep significantly, past the initial truncation radius. This timescale depends on the viscosity of the gaseous disk (Appendix \ref{app:disk_evolution}).

\textbf{Phase III - Resonance Sweeping: $t \gtrsim t_{\rm viscous}$}

The sweeping of the secular resonance relies on the evolution of the depleting gas disk. Thus, this phase starts at a timescale comparable to the viscous timescale of the disk. The position of the secular resonance depends on the relative masses of the inner and outer disk ($M_{\rm in}$ and $M_{\rm out}$, respectively) at a given time,  compared to the mass of the giant, as described in  \hyperlink{cite.Best2024}{Paper I}:
\begin{eqnarray}
\left( \frac{a_{\rm res}(t)}{a_{J}} \right) ^{3/2} = \frac{ 0.012 M_{\rm out}(t) + 3.83 M_{\rm in}(t) }{M_{J}}.
\end{eqnarray}
The resonance sweeps inward, as described in \hyperlink{cite.Best2024}{Paper I}, while the gas disk depletes. However, the key difference with our previous results lies in the underlying size distribution. By this stage, most of the solid material has either already been ground down to dust or has formed large planetesimals (radius > 1000 km), which are removed from the simulation. 
The remaining material continues to grind down due to the high eccentricities induced by the inward-moving resonance. However, as we show in Section \ref{sec:mat_trans_fid}, the large planetesimals that were taken out of the simulation undergo very efficient transport, possibly with little fragmentation. 

Phase III concludes once the gaseous disk is fully depleted. 
While we end our simulations
at 10 Myrs, in a more realistic scenario, the process would end when photoevaporation from the star dissipates the inner gas disk, effectively shutting down any remaining accretion and transport mechanisms---an effect that is not included in our simulations.

\subsection{Transport of material} \label{sec:mat_trans_fid}

We aim to address whether material transport remains efficient when collisions among planetesimals are taken into account. In \hyperlink{cite.Best2024}{Paper I}, we observed that the transport ranged from 10 - 80\% of the initial solid reservoir, however the impact of collisions on this process remained an open question.

Our results indicate that most collisions occur before the resonance sweeps through the inner disk. At this stage, once the resonance starts sweeping inwards, the planetesimals are already divided into two populations. A small population (less than 1\% contribution to the total mass budget, see phase III in Figure \ref{fig:fig2}) with sizes ranging from 1 km to $\sim$10 km, and a larger population of massive planetesimals (above 1000 km in size) representing around 50-60\% of the original solid mass reservoir.

\begin{figure}
    \centering
    \includegraphics[width = 1.1 \linewidth]{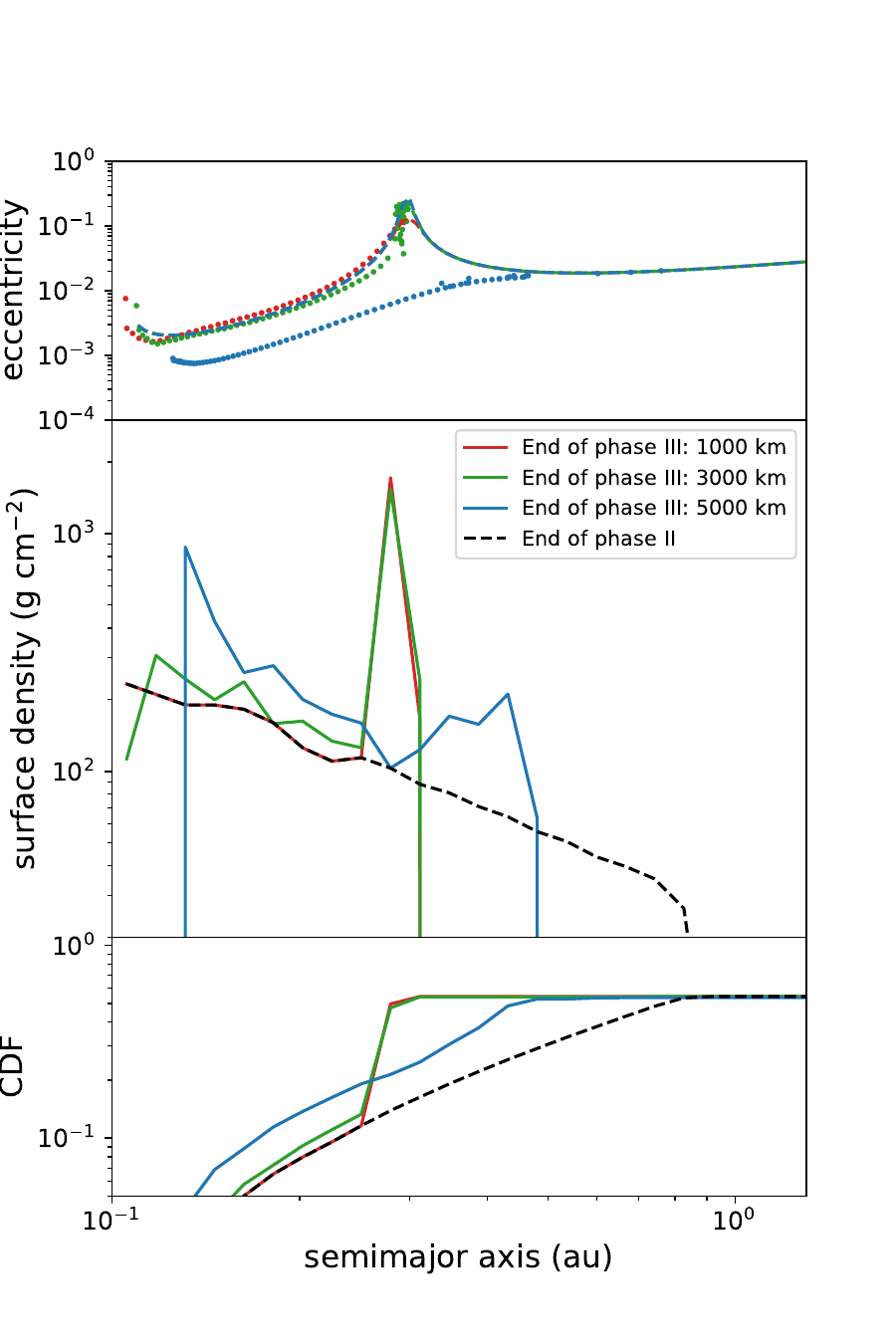}
    \caption{Eccentricity (top) and solid surface density (middle) as a function of semimajor axis for planetesimals of 1000, 3000 and 5000 kms in red, green and blue, respectively, affected by the secular resonance and the Lindblad torque, not considering collisions. The bottom panel shows the cumulative distribution function (CDF) for the amount of mass as a function of semimajor axis, normalized by the initial amount of solid material in the simulation ($\approx$ 9 $M_\Earth$). The dashed lines in the upper panel correspond to the forced value without considering the Lindblad torque, and dashed lines in the middle and lower panels correspond to the solid mass distribution at the end of phase II.}
    \label{fig:3}
\end{figure}

Thus, we focus on how the sweeping secular resonance influences the population of large planetesimals formed early in phase I. To investigate this, we conducted tests involving the migration of these planetesimals. Since their size distribution remains uncertain as these planetesimals 
would continue to grow after being removed from the simulation, we performed a set of separate simulations with the same total solid mass (corresponding to the material that exceeded our cutoff size of 1000 km), assuming fixed planetesimal sizes of 1000, 3000, and 5000 km to assess the efficiency of their transport. 

Due to the large size of these planetesimals, we included the effects due to Lindblad torques for both semi-major and eccentricity in addition to the dynamical effects described in Section~\ref{sec:methods} (see Eqs. \ref{eq:1}--\ref{eq:3}). According to the linear theory by \cite{tanaka_2002}, these operate on timescales of:
\begin{equation}
    \frac{e}{|\dot{e}|} = \tau_{e, {\rm Lindblad}} = 0.13 \frac{M_\star}{m_p} \frac{M_{\star}}{\Sigma_{\rm gas} a_p^2} 
    h_{\rm gas}^4 \left[ 1 + \frac{1}{4} \left( \frac{e_p}{h_{\rm gas}}\right)^3 \right] \Omega_p^{-1}, \nonumber
    \label{eq:tau_e_lindblad}
\end{equation}
\begin{eqnarray}
    &\approx& 0.1 \hspace{0.1cm}{\rm Myrs} \left( \frac{a_p}{1 \rm au} \right)^{1/2}  \left( \frac{R_p}{1000 \rm km} \right)^{-3} \left( \frac{\Sigma_{\rm gas}}{10^3 \rm g/cm^2} \right)^{-1} \times \nonumber \\ &&\left( \frac{\rho_{\rm solids}}{3 \rm g/cm^3} \right)^{-1} \left( \frac{M_*}{M_{\Sun}} \right)^{3/2}
\end{eqnarray}
and
\begin{equation}
    \frac{a}{|\dot{a}|} = \tau_{\rm a, Lindblad} = 0.26 \frac{M_*}{m_p} \frac{M_*}{\Sigma_{\rm gas} a_p^2} h_{\rm gas}^2 \Omega_p^{-1}    ,
\end{equation} 
where $\Omega_p = \sqrt{G (M_{\star}+m_p) / a_p^3}$ is the mean motion of the planetesimal, $m_p$ its mass (assuming a 1000 km planetesimal with a density of 3 g/cm$^2$) and $h_{\rm gas}$ its local aspect ratio which at 1 au is roughly 0.04 according to our model. Note that $h_{\rm gas} \propto a_p^{1/4}$ due to the disk's temperature profile (\hyperlink{cite.Best2024}{Paper I}), and the second line in Equation (\ref{eq:tau_e_lindblad}) assumes $e_p\ll h_{\rm gas}$.

In Figure \ref{fig:3}, we show the results of these simulations, where we observe a significant amount of transported material for planetesimals up to 3000 km. In this case, material is effectively swept up and concentrated into a narrow ring along the path of the secular resonance. For the 1000 and 3000 km populations, we get a transport efficiency of $\sim 40\%$ at the peak, as illustrated in the lower panel of Figure \ref{fig:3}, which shows the cumulative mass distribution as a function of semimajor axis. This distribution is normalized to the initial amount of solid material at t=0 of the simulation (corresponding to $\approx 9 M_\Earth$). In contrast, the 5000 km population decouples from the secular resonance due to the lower eccentricities achieved, resulting in a reduced transport efficiency. In this case, sweeping is less efficient, only $\sim 25\%$ of material is transported at the peak, and the distribution is more spatially extended.

The decoupling of the 5000 km planetesimals from the secular resonance is primarily due to their large size and the enhanced eccentricity damping due to the Lindblad torques. We estimate that the Lindblad torque becomes significant for planetesimals above 5000 km at 1 au, compared to the secular timescale:
\begin{eqnarray}
    \tau_{e, {\rm secular}} &=& \frac{e_p}{|A_{\rm pJ}|e_{J}}   \\ \nonumber
    &\approx& 0.8 \hspace{0.1cm}{\rm kyrs} \left( \frac{a_p}{1 \rm au} \right)^{-5/2} \left( \frac{M_J}{3 M_{\rm Jup}} \right)^{-1} \left( \frac{e_J}{0.05} \right)^{-1} \\ \nonumber
    &\times& \left( \frac{a_J}{3 \rm au} \right)^{4} \left( \frac{M_*}{M_{\odot}} \right)^{1/2} \left( \frac{e_p}{10^{-2}} \right),
\end{eqnarray}
where $A_{\rm pJ}$ characterizes the eccentricity excitation due to the planet (given by Eq. 21 in \hyperlink{cite.Best2024}{Paper I}).

\subsection{Varying parameters} \label{sec:varying_params_result_S}

Having demonstrated that a significant fraction of the initial solid material reservoir is converted into planetesimals larger than 1000 km in radius in our fiducial case, we now examine whether this outcome is robust across a broad parameter space by varying one parameter at a time in a series of tests. Some variations correspond to different physical scenarios, such as changes in the mass or eccentricity of the Jupiter, which we explore in the next section. Other tests are aimed at evaluating our modeling assumptions. For instance, by varying the initial distribution of solid material (initial amount of solids or initial size of the planetesimals). Finally, we modify parameters related to the resolution (e.g. number of rings, mass bins, time step limit coefficients) to assess whether the resolution adopted for the fiducial case (constrained by computational cost) produces representative results.

We present here the tests conducted on giant planet properties and defer the discussion of other tests to Appendix \ref{app:resolution}, where we show that our results are not significantly affected by changes in resolution parameters and modeling assumptions.

\subsubsection{Giant planet parameters} \label{sec:giant_planet}

\begin{figure*}[]\center \label{fig:cdf}
\includegraphics[width=18cm]{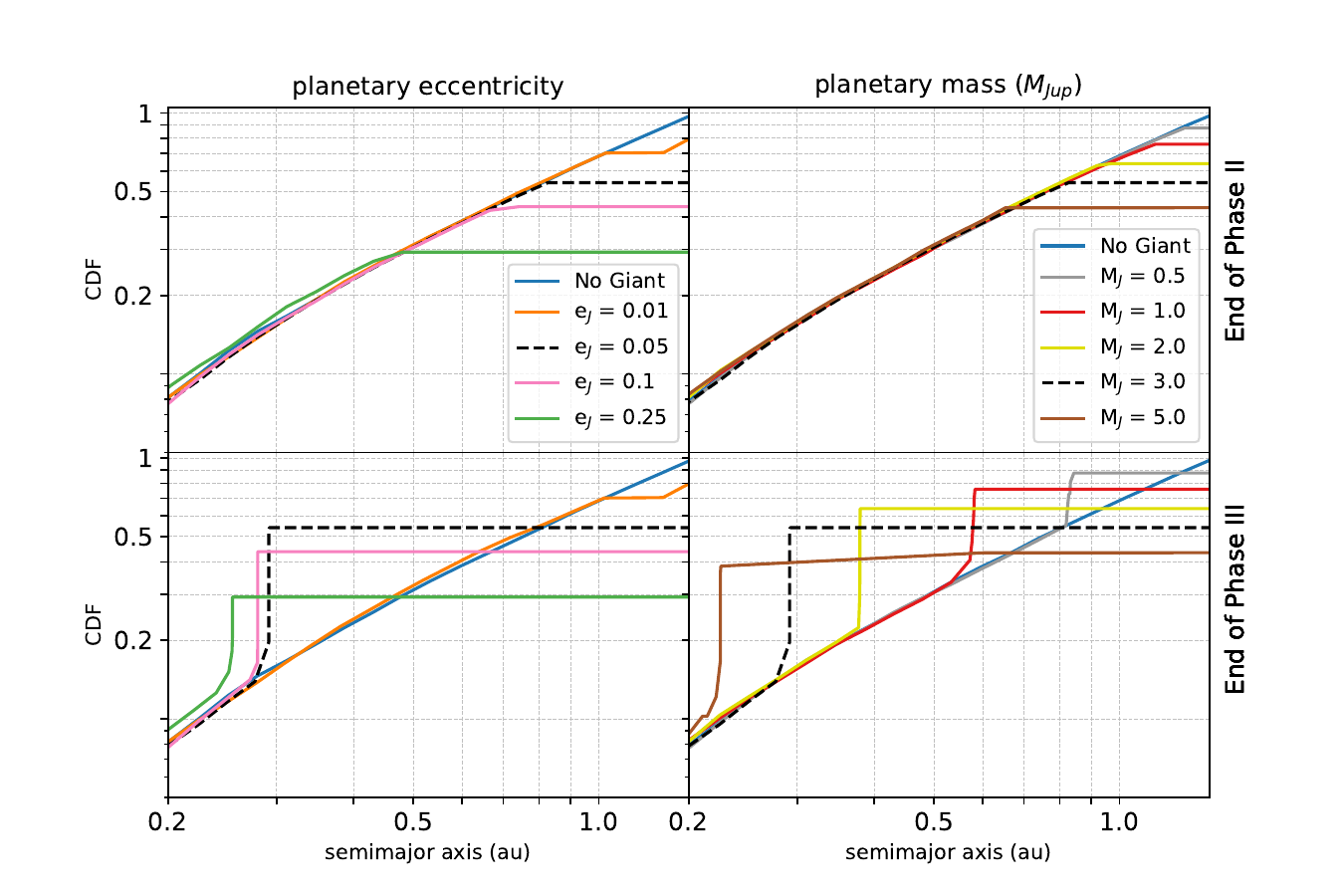}
\caption{Cumulative distribution function of total planetesimal mass as a function of the semi-major axis for various eccentricities (left panels) and masses (right panels) of the giant planet. Upper panels correspond to the end of phase II, with no material having been swept by the secular resonance, and lower panels correspond to the end of phase III where the resonance has swept and accumulated material within the disk's inner parts.}
\end{figure*}

We ran a set of simulations for the migration of planetesimals influenced by the sweeping secular resonance, taking as the initial distribution of solids the planetesimals larger than 1000 km at the end of phase II for the fiducial case. The results are presented in Figure~\ref{fig:cdf}, where we present the cumulative distribution function for 1000 km planetesimals under variations in the giant planet's eccentricity (left) and mass (right). The upper panels correspond to the end of phase II, when all large planetesimals have formed, but before the secular resonance begins to sweep. The bottom panels depict the end of phase III, after the resonance has swept the material. Note that not all CDFs reach 1 due to mass loss from fragmentation and the imposed lower limit on mass.

When the planetary eccentricity is varied, the secular resonance location remains the same, since the underlying theoretical framework is linear when it comes to planet-induced perturbations (see Section \ref{sec:methods} and \hyperlink{cite.Best2024}{Paper I}). However, looking at the left column of Figure \ref{fig:cdf}, it is evident that increasing the planetary eccentricity affects the CDF of planetesimals. This is because higher $e_J$ leads to higher planetesimal eccentricities, which, in turn, affect the truncation radius (see Eq.~\ref{eq:e_frag}). By the end of phase III (bottom left panel), after the resonance has swept through the disk, we observe that higher planetary eccentricities result in a more inwardly shifted accumulation of material. The CDFs follow a power-law distribution up to this point, beyond which no material remains.

A similar comparison can be made for the simulations in which the planet’s mass is varied (right panels). We find that the location of the secular resonance depends on the mass of the giant planet (see \hyperlink{cite.Best2024}{Paper I}), which influences the truncation radius through the forced eccentricities of planetesimals. Since the disk mass is held constant, a lower mass giant means the initial resonance position is closer to the planet. This also influences the outcome seen at the end of phase III, where the accumulation of material shifts outward in the lower mass case. As in the eccentricity tests, the final distributions follow a power-law up to the position where material is accumulated and drop off beyond.

In both variational cases, we observe a sharp increase just before the region where material has been depleted at the end of phase III. This jump corresponds to the accumulated material by the secular resonance, assuming all planetesimals are 1000 km. As shown in Figure \ref{fig:3}, planetesimals of 3000 km exhibit very similar behavior to the 1000 km ones, while 5000 km planetesimals are not efficiently transported due to the Lindblad torque dominating, which inhibits their coupling to the resonance.

The only cases where material is not accumulated correspond to the no planet case in Figure \ref{fig:cdf}. In all other cases, when comparing the amount of material in the jump in lower panels (which corresponds to the peaks in accumulated material), we find values ranging from $\sim$15 - 40\%. This represents an enhancement of about 2 to 4 times the case without a giant. The highest amount of transported material, around 40\%, is achieved with planetary masses between 1-3 $M_{\rm Jup}$. In contrast, for the higher eccentricities, this amount decreases rapidly to 15\% for values exceeding the fiducial eccentricities ($e_J = 0.05$). Although one might expect higher eccentricities to transport material more efficiently, these cases contain less material by the end of phase II, as their truncation radius is already too close to the inner edge of the disk.

The amount of transported mass easily scalable, as our tests indicate that increasing the initial amount of solid material above 5\% of the disk gas mass yields an almost constant fraction of solids being converted into planetesimals larger than 1000 km (see panel A in Figure \ref{fig:parameters_giant}).

\subsubsection{Other simulations} \label{sec:other_simulations}

In all previous simulations we have assumed that the giant planet is fully formed at $t=0$. Although formation is expected to occur  very early on \citep{guilera2020}, it is worth considering the effects during the planet's growth phase. To this end, we ran a simulation where the Jupiter grows exponentially with a characteristic timescale of $10^{5}$ yrs. In this simulation, the secular resonance starts closer to the planet and only reaches the starting position of the fiducial case at times comparable to the growth timescale. Consequently,  formation of large planetesimals happens in all the region considered (0.1 -- 1.5 au) reaching close to 99\% efficiency in converting initial solid material into 1000 km planetesimals.

Another consideration is the choice of the parameter $\sigma_e$. As we consider a ratio of 2 to 1 for $\sigma_e$ to $\sigma_i$, this also affects the inclination distribution of planetesimals, making them more concentrated in the midplane for lower values of $\sigma_e$.

As in the work by \cite{silsbee}, we did not add a dependence on $\sigma_e$ for different sizes or times. Such dependencies may emerge from the fact that larger planetesimals decouple from the gas and take more time to settle to the forced eccentricity value. However, the growth in our simulations is limited by smaller sizes grinding down, and these planetesimals (1 -- 10 km) settle quickly ($\sim 1 ~\rm kyr$, see also \hyperlink{cite.Best2024}{Paper I}). We ran an additional simulation using $\sigma_e=10^{-3}$ (instead of the fiducial value of $\sigma_e=2\times10^{-4}$) and found that in this case, planetesimals are unable to grow above $\sim$ 100 km. Thus, the choice of $\sigma_e$ is crucial for the formation of large planetesimals.

Finally, we explore implications of lowering the dust-to-gas ratio from 5\% to 1\%. The fiducial choice of a 5\% solid ratio was motivated by the amount of solids required to form Kepler-type systems \citep{extrasolarnebula}. However, a 5\% solid ratio might be in the upper end, considering that solids might be filtered by the giant, potentially reducing their value inherited from the ISM. We ran a simulation where we considered 1\% solid ratio, while simultaneously increasing the amount of gas mass to 5\% solar mass. Thus, the total amount of solids remains the same as the fiducial case. We found that this scenario is more efficient in forming large planetesimals, achieving an efficiency of 85.5\% compared to 53.8\% in the fiducial. This increased efficiency results from the more massive disk causing the secular resonance to start closer to the giant, thereby moving the truncation radius further out.

\section{Discussion} \label{sec:discussion}

We now discuss the implications of our findings for planet formation, their relation to observed trends in planetary systems, and the limitations of our work, along with possible improvements.

\subsection{Consequences for inner planet formation}

We first detail the predictions of our model 
in relation to the
observational trends 
discussed in Section \ref{sec:introduction}.

\paragraph{Super-Earth formation from rings} Throughout this work, we have shown that our model consistently produces an overdense ring of moon-sized objects.  From Figure \ref{fig:cdf} we find that the rings are located at $\sim 0.1-0.5$ au and have narrow widths ($\Delta a/a\lesssim 0.1$). 
Previous studies have shown that overdense rings, when simulated using N-body codes, can give rise to a system of super Earths in the case of extra solar systems \citep{batygin_2023}, and the terrestrial planets in the case of our Solar System \citep{hansen_2009}. Our findings may provide the initial conditions or seed populations needed for such studies. 

\paragraph{Giant planet and Super-Earth correlation} In our models, the total amount of solid mass in large planetesimals (>1000 km) is similar both with and without a giant planet. However, the amount of accumulated material is stronger for a giant with higher eccentricity, which may explain the correlation found in \cite{friends_not_foes}. At this stage of our work, we do not predict a clear over- or under-abundance of super-Earths in systems with a giant planet. 
Nevertheless, the presence of a giant planet might imply a higher amount of solid mass in the protoplanetary disk, which might lead to the formation of more large planetesimals and, thereby, increase the likelihood of super-Earth formation, assisted by the sweeping of a secular resonance described here.

\paragraph{Peas-in-a-pod and Spacing complexity} The secular resonance induced by a sufficiently eccentric giant planet and a massive gaseous disk redistributes solid material within the disk by the sweeping mechanism, in contrast to a system without a planet (or a giant low in mass and/or eccentricity), as can be seen from the bottom panels in Figure \ref{fig:cdf}. Without a giant planet, the formation of large planetesimals in the inner disk follows a standard power-law size distribution. However, when we incorporate a giant, the secular resonance greatly modifies the distribution of solid material, potentially explaining observed patterns in the spacing and gap complexity of inner planetary systems \citep[see, e.g.,][]{heweiss}. A case study using Kepler-139 is provided in Doty et al (in prep.).

\bigskip
In closing, although the total mass in large Moon-sized planetesimals may remain similar with or without a giant planet, the spatial redistribution of this material, driven by a sweeping secular resonance, can significantly alter the architecture of the inner system. Thus, while we do not currently predict a clear abundance trend in super-Earths, the presence of a giant planet may still influence their spacing, size distribution, or the formation of gaps.

\subsection{The critical role of disk gravity}\label{sec:ndg}

Previous related studies have ignored the gravitational effects of the gaseous disk on the apsidal precession rates of the planetesimals \citep[e.g.,][]{kortenkamp,Guo2023}. In the absence of disk gravity, the eccentricity profile of planetesimals $e_p(a_p)$ differs significantly. First, instead of peaking at a specific location interior to the giant planet (the secular resonance), $e_p(a_p)$ increases linearly towards the cold Jupiter (i.e., $e_p \propto a_p/a_J$; \citealt{Murray1999}). 
Second, the magnitude of $e_p(a_p)$, especially at large separations from the planet, would be larger when the disk's gravity is ignored \citep[see also][]{Sefilian+2021, sefilian_stirring}.  These effects combined significantly shift the truncation radius inward, reducing the amount of material available for subsequent phases of planetary formation (as shown in panel C in Figure  \ref{fig:parameters_giant}).

Additionally, since the exterior planet remains in a fixed position in our simulations, this eccentricity profile has no time dependence. In our model, time dependence in the eccentricity profile arises from the depletion of the gas component. The absence of this effect means there is no sweeping mechanism, a crucial factor for material transport, as highlighted here and in \hyperlink{cite.Best2024}{Paper I}. Moreover, as shown in our previous work, material transported in a scenario without disk gravity is minimal and consists mostly of the smaller planetesimals subject to the gas drag (1-10 km). Consequently, when the majority of the solid mass is concentrated in larger planetesimals, the amount of transported material becomes negligible.

These findings point towards the critical role of disk gravity in shaping the dynamical and collisional evolution of planetesimals. Incorporating disk gravity in the simulations is, therefore, essential for accurately capturing the material growth and redistribution during planet formation.

\subsection{Caveats and Future Work}\label{sec:caveats}

\subsubsection{Effects of Self-stirring and Scattering} \label{sec:scattering}

Large planetesimals (with radii larger than 1000 km) begin to form in our simulations within 0.1–1 kyr (Phase I). Once formed, they are removed from the simulation, as their continued presence would require accounting for scattering effects, which are not included in our model. Additionally, due to their substantial size, these planetesimals are strongly resistant to fragmentation, as their self-gravity holds them together. As a result, most collisions involving such bodies lead to mergers rather than further fragmentation, preventing the regeneration of smaller planetesimals that could re-enter the simulation.

While there may be concerns that these large planetesimals could influence the smaller ones still present in the simulation, we argue that the timescales for tidal eccentricity damping of smaller planetesimals by larger ones are too long to have any significant impact. As shown in Figure 3 of \cite{tidal_timescale}, the damping timescales for 1000 km sized planetesimals is on the order of 1 Myr and even longer for smaller planetesimals. By this time, our simulations have already entered phase III, where the remaining solid material in the disk is negligible (less than 0.1 Earth masses).

Therefore, removing large planetesimals from the simulation has a negligible impact on the results, as their dynamical influence on the remaining planetesimal population occurs on much longer timescales than the relevant phases of our simulations.

\subsubsection{Lifetime of the disk and end of simulations}

Phase III formally ends when the gaseous disk has fully dissipated, as this marks the end of dynamical interactions between gas and planetesimals, effectively halting secular resonance-driven migration. In our model, we approximate this by stopping the simulation at 10 Myr, though in reality, a disk's lifetime may vary from 1 to 10 Myrs \citep{lifetime_disk}.

After 10 Myrs the disk is most likely dissipated in most cases. However, our simulations do not fully capture this process, as they omit key physical mechanisms driving disk evolution, such as internal photoevaporation. Inside-out photoevaporation, triggered by high energy radiation from the central star, is known to play a crucial role in clearing the inner regions of gaseous disks and accelerating their dissipation \citep{picogna}. By neglecting this effect, our model may overestimate the disk's lifetime, potentially extending artificially the duration of phase III beyond what would occur in reality.

Future iterations of our model should incorporate these disk evolution processes to provide a more realistic timeline for the planetesimal evolution and their interaction with the disk environment. This will enable a more accurate characterization of the transition from the protoplanetary disk phase to the final architecture of the planetary system.

\subsubsection{Physical properties of planetesimals}

Our simulations demonstrate that large planetesimals (exceeding 1000 km in radius) can form and be efficiently transported via the combined effects of secular resonance and gas drag. This process results in a substantial redistribution of the available solid material within the disk.

The next step in this research will focus on incorporating a detailed treatment of collisions and evolution among these large planetesimals which were taken out of the simulation and model them, as well as their effects on nearby planetesimals. This requires accounting for their impact velocities and applying the merger/fragmentation criteria previously established in our model. Given the relatively small number of massive planetesimals that form during the early stages of evolution (within the first 10 kyr), we propose using an N-body integrator to simulate their interactions. This approach will allow us to accurately track their dynamics, assess growth through mergers, and model fragmentation outcomes for individual collision events. We anticipate this final step will bring us closer to understanding the final architecture of the system, and provide deeper insights into the observed correlations in planetary systems.

Another possible point of improvement to our model would be to incorporate more realistic physics in the collisions, for example, here we only considered planetesimals made entirely out of strong basalt, however, if we include material from different parts of the disk, some may have ices which would change their density as well as their resistance to fragmentation.

\section{Conclusion}\label{sec:conclusions}

Building on the sweeping secular resonance mechanism explored in \hyperlink{cite.Best2024}{Paper I}, we show that planetesimal collisions do not disrupt material transport. Instead, this redistribution of solids, driven by the presence of a giant companion and an evolving gaseous disk, occurs on a longer timescale than planetesimal growth. This allows large, Moon-sized bodies (>1000 km) to first form via collisions and then accumulate in narrow rings near its inner edge. Our simulations yield an accumulation of about 5 Earth masses (about 60\% of the initial solid material in the disk) in a small compact ring ($\Delta a/a\lesssim 0.1$) centered at around 0.1 and 0.5 au.

By including collisions, we find that a giant companion with $e \gtrsim 0.01$ and $\gtrsim$ 1 $M_{\rm Jup}$ enhances the amount of solid material near the inner edge of the disk by a factor of 2-4 compared to systems without such a planet. Disk gravity also proves essential: omitting it leads to fewer large planetesimals being formed due to their higher forced eccentricity and also to a less efficient sweeping transport mechanism, as identified in \hyperlink{cite.Best2024}{Paper I}.

This process has significant implications for the early stages of planet formation and may lead to planetary architectures different than the commonly observed peas-in-a-pod systems, where sibling planets exhibit similar masses and orbital spacings. Instead, the enhanced concentration of large planetesimals in narrow regions could favor the formation of planetary cores in localized areas. 

Although we do not model the subsequent accretion of these $\sim$1000 km bodies into fully formed planets, our results provide the groundwork for future studies of how their transport, composition, and spatial distribution shape the emergence and diversity of planetary systems.

\acknowledgements
M.B., C.C., and C.P. gratefully acknowledge support from the John and A-Lan Reynolds Faculty Research Fund.
M.B. acknowledges additional support from the National Agency for Research and Development (ANID) / Scholarship Program / Doctorado Nacional grant 2021 - 21211921,  CASSACA grant CCJRF2105, by the ANID BASAL Center for Astrophysics and Associated Technologies (CATA) through grant FB210003 and from the ANID -- Millennium Science Initiative Program -- Center Code NCN2024\_001.
A.A.S. is supported by the Heising-Simons Foundation through a 51 Pegasi b Fellowship.
C.C. is supported by ANID through FONDECYT grant n$^\circ$3230283. 
C.P. acknowledges support from ANID Millennium Science Initiative-ICN12\_009, CATA-Basal AFB-170002, ANID BASAL project FB210003, FONDECYT Regular grant 1210425, CASSACA grant CCJRF2105, and ANID+REC Convocatoria Nacional subvencion a la instalacion en la Academia convocatoria 2020 PAI77200076.

\appendix

\renewcommand{\thefigure}{A\arabic{figure}}
\setcounter{figure}{0}

\section{Gaseous disk evolution module} \label{app:disk_evolution}

In \hyperlink{cite.Best2024}{Paper I}, we developed a code for modeling the viscous evolution of the gaseous disk, which is now available on GitHub.\footnote{\url{https://github.com/marcybest/diskevolution}}
By default (as in this work), the module initializes the protoplanetary disk using a power-law surface density profile with index $\gamma$, incorporating an exponential taper beyond an outer cutoff radius $r_{\rm cut}$:
\begin{equation} \label{eq:initial_profile}
    \Sigma_{\rm gas}(r) =  \Sigma_{\rm gas,0} \left(\frac{r}{1 \rm au}\right) ^ {-\gamma} \exp\left[-(r/r_{\rm cut})^{2-\gamma}\right].
\end{equation}
The user can also define their own initial density function. We then use a full implicit Crank-Nicholson method to evolve this 1-D profile by solving the viscous evolution equation under the assumption of  zero torques at the boundaries. We update the gas surface density profile $\Sigma_{\rm gas}$ using the standard diffusion equation \citep{Pringle1981}:
\begin{eqnarray}\label{eq:viscosity}
    \frac{\partial\Sigma_{\rm gas}}{\partial t} &=& \frac{1}{r} \frac{\partial}{\partial r} \left[ 3r^{1/2} \frac{\partial}{\partial r}(\nu \Sigma_{\rm gas} r^{1/2}) \right] ,
\end{eqnarray}
where $\nu = \alpha c_s H$, $c_s$ is the sound speed, and $H$ is the scale height. We then account for a gap profile produced by the giant planet using a template from \cite{Duffell_2020} and multiplying it with the gas surface density function to mimic the effects of the giant. For more details on how this module works and the assumptions made when selecting the different parameters, see \hyperlink{cite.Best2024}{Paper I}.

\begin{figure} \label{fig:A1}
    \centering
    \includegraphics[width=10cm]{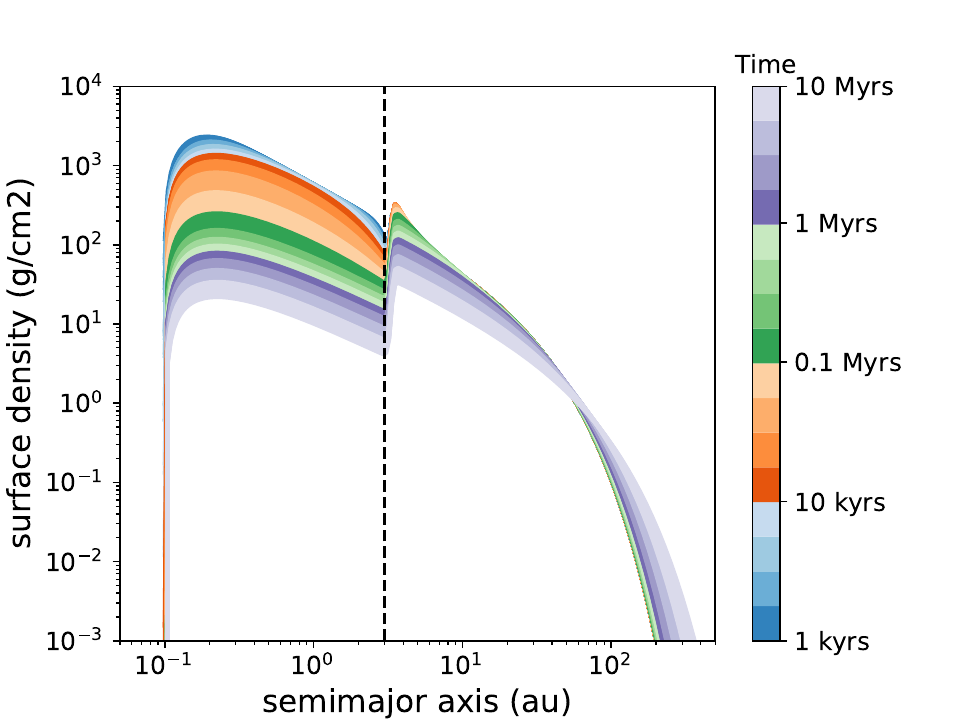}
    \caption{Temporal evolution of the surface density of the gaseous disk for our fiducial  two alpha model ($\alpha_{\rm in}=10^{-3}, \alpha_{\rm out}=10^{-4}$). The position of the giant is marked with a dashed black line. } 
    
\end{figure}

With this, we model a gaseous protoplanetary disk that evolves purely through viscous processes, without considering additional physical effects such as interactions with giant planets, photoevaporation winds, etc. In our fiducial calculations, two different $\alpha$ viscosity values are considered: one for the inner region and another for the outer region beyond the giant planet’s orbit ($a_J = 3$ au). The results of such a calculation are shown in Figure \ref{fig:A1} assuming $\gamma = 1$ and $r_{\rm cut} = 
25$ au. Note that the viscosity contrast leads to a transient accumulation of material at the transition point, creating a "traffic jam" effect due to the differing gas migration rates on either side.

\renewcommand{\thefigure}{B\arabic{figure}}
\setcounter{figure}{0}

\section{Resolution tests}\label{app:resolution}

\begin{figure*}\label{fig:parameters_resolution}
    \includegraphics[width=1\textwidth]{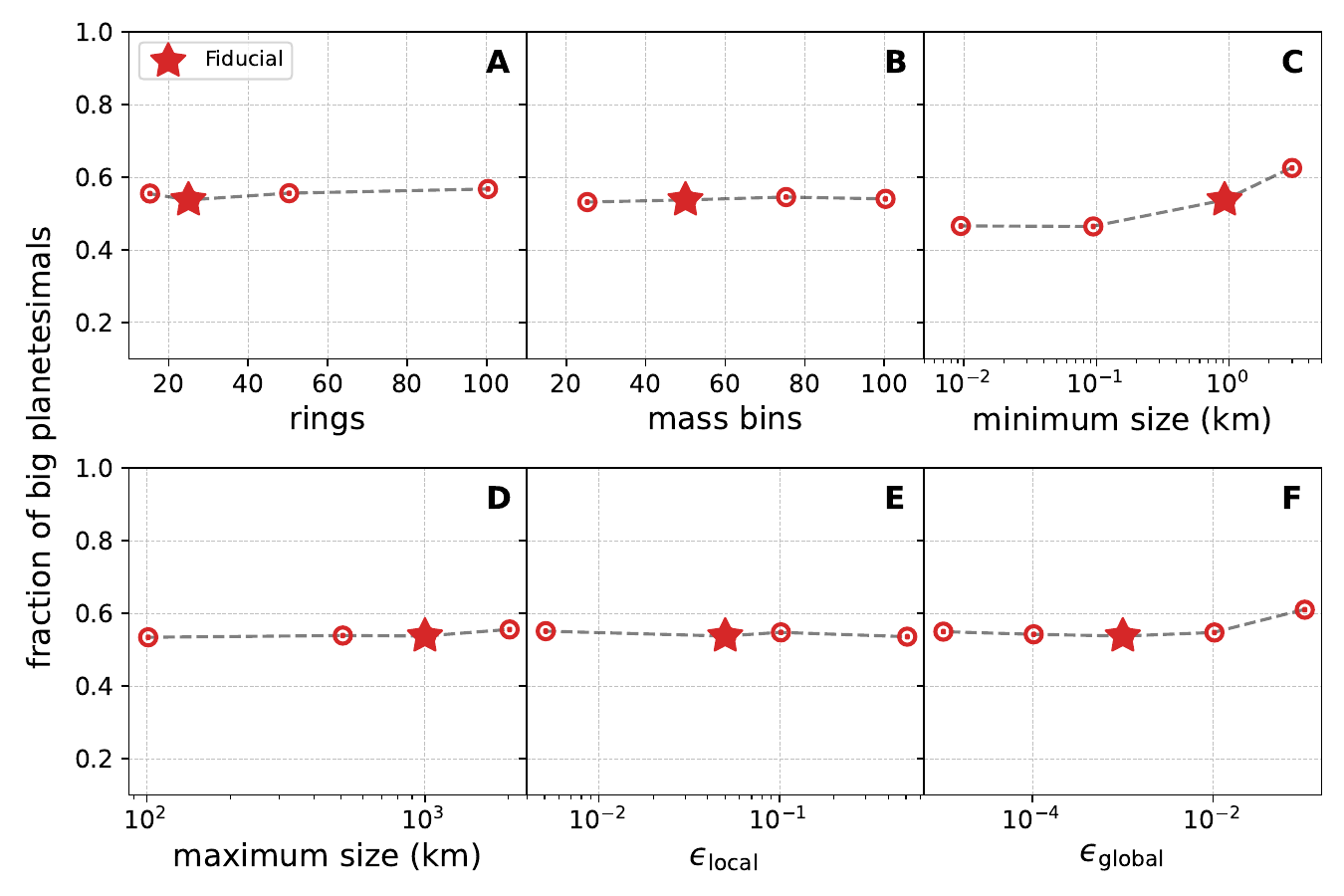}
    \caption{
    Impact of varying resolution parameters on the efficiency of planetesimal formation. Specifically, we display the fraction of solid material converted into planetesimals larger than 1000 km for each test case. Each panel corresponds to a different resolution parameter being varied, while keeping all others fixed. The fiducial case is included in all panels for reference and is marked with a star.}
\end{figure*}

We conducted simulations to test the impact of the resolution parameters on the collisional module, including the number of annuli, number of mass bins, upper and lower mass bins, $\epsilon_{\rm local}$, and $\epsilon_{\rm global}$. The corresponding results are summarized in Figure \ref{fig:parameters_resolution}. These tests ensure that our fiducial case is sufficiently resolved to capture the relevant collisional activity without excessive computational cost. We find that, in most cases, varying these parameters away from the fiducial values does not lead to significant differences. However, an exception is $\epsilon_{\rm global}$, for which values above 0.01 seem to yield different results. This parameter is associated with the calculation of time steps, and setting it too high reduces constraints in the amount of mass change within a given bin. Another parameter impacting our results is the minimum size of planetesimals, with outcomes stabilizing below 100 m. Although this effect is not significant, it should be taken into account for future tests.

\section{Testing the collisional code} \label{app:tests}

Appendix A of \cite{silsbee} outlines a series of tests that can be used to validate the new collisional module in our code. Since our modifications introduce coagulation and fragmentation effects, it is essential to test these processes separately to ensure their proper implementation.

\subsection{Coagulation Test}
\renewcommand{\thefigure}{C\arabic{figure}}
\setcounter{figure}{0}

\begin{figure} \label{fig:coagulation}
    \centering
    \includegraphics[width=0.5\linewidth]{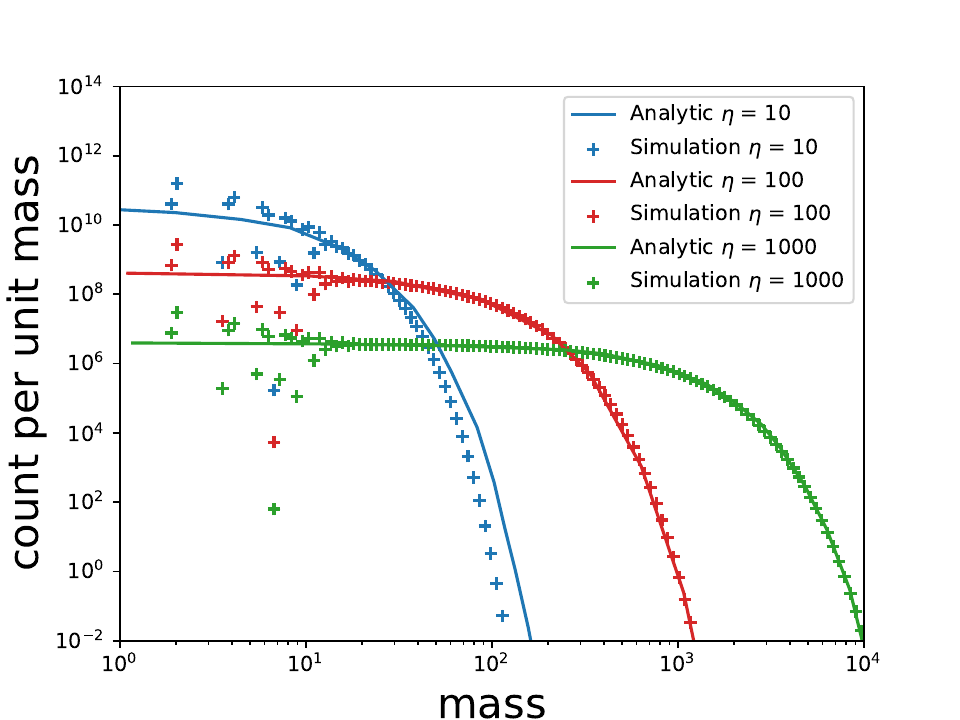}
    \caption{Coagulation test. Crosses represent number of particles ($N$) per unit mass as a function of mass at three different snapshots, expressed in terms of $\eta=Nt$, only considering coagulation (i.e., all mergers, no fragmentation). Solid lines show the analytical solution from \cite{smoluchowski} for each corresponding value of $\eta=Nt$ (see the legend).}
    \label{fig:enter-label}
\end{figure}

Turning off fragmentation allows us to isolate and test the coagulation module in our code. In this simplified scenario, planetesimals grow exclusively through perfect mergers in every collision, ensuring that no mass is lost during interactions. This setup begins with a uniform population of identical planetesimals, each assigned a dimensionless mass of 1, and then tracks their growth as they merge to form larger bodies. This setup follows the classical framework described in \cite{smoluchowski}. In Figure \ref{fig:coagulation}, we compare our results with the analytical solution for the mass spectrum finding a good agreement at larger masses, as smaller ones are dominated by noise. In this case, the collision rate is determined by the following relation:
\begin{equation}
    \frac{dn_k}{dt} = \frac{1}{2}  \sum_{i+j = k} n_i n_j - n_k \sum_{i} n_i.
\end{equation}

We find that the coagulation module reproduces results in agreement with theoretical expectations from \citet{smoluchowski}.

\subsection{Fragmentation test} \label{app:fragmentation_test}

We now test the fragmentational cascade module. As each collision produces a tail of fragments with a specific distribution, we have to make sure this corresponds to known results. Although the temporal evolution of the distribution is not known, steady state solutions are well known \citep{obrien2013}. In the steady state, the logarithmic slope in the number of planetesimals as a function of size, $d\ln N/d\ln R = -z$, should be consistent with known solutions
\begin{equation} \label{eq:index}
    z = \frac{3 + \xi}{6 + s},
\end{equation}
where $\xi$ is the scaling factor for the cross section of the planetesimals $A_g \propto 
R^{\xi}$ and s is the scaling factor for $Q_{\rm *} \propto R^{s}$. For this test, we set $\xi = 2$ and $s=0$.

\begin{figure}
    \centering
    \includegraphics[width=0.75\linewidth]{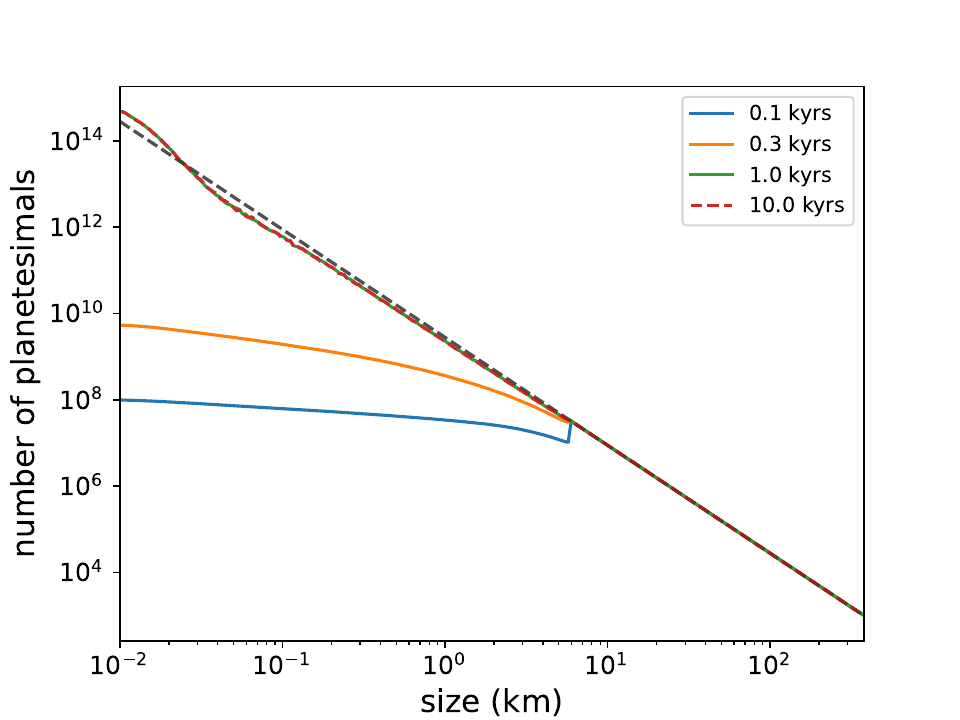}
    \caption{Fragmentation test. Number of planetesimals of each size as a function of size. The dashed line corresponds to the analytical value our results should approach, in this case a power-law with slope -5/6 (see Eq. \ref{eq:slope_v}).}
    \label{fig:test_fragmentation}
\end{figure}

Figure \ref{fig:test_fragmentation} shows that the size spectrum gradually evolves toward a power-law distribution with index $-5/6$ described in Eq. (\ref{eq:index}). At smaller sizes, a wave-like pattern emerges, as already described by \cite{obrien2013} due to the lower limit in mass. Since the cascade extends to indefinitely small sizes, some mass is lost at this boundary. To prevent mass depletion over time, we maintain a fixed number of planetesimals in the upper 40\% of size bins throughout the simulation, ensuring their population remains unchanged.

\subsection{Equilibrium in mass spectrum} \label{app:equilibrium}

\begin{figure*}\label{fig:Qstar}
    \includegraphics[width=0.5\textwidth]{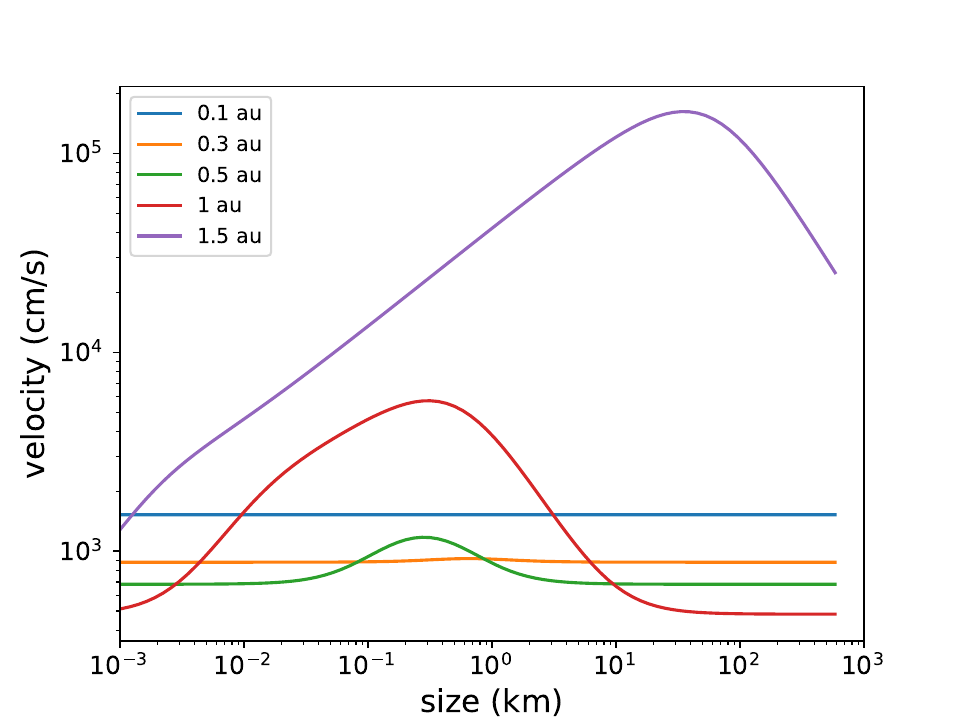}
    \includegraphics[width=0.5\textwidth]{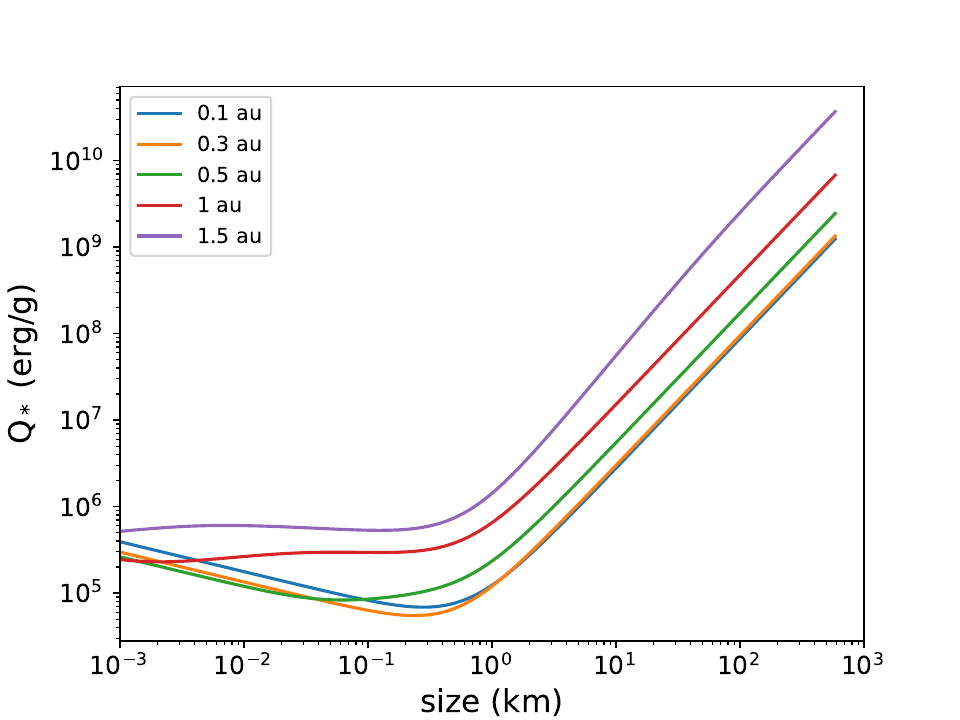}
    \caption{Encounter velocity (left) and fragmentational threshold energy $Q_*$ (right) as a function of particle size for the fiducial case at t=0 in our simulation. For the right panel we considered two equal sized particles, while for the left panel they correspond to particles of size R and 2R.}
\end{figure*}

As discussed in Appendix \ref{app:fragmentation_test}, the mass spectrum reaches a steady-state solution, which generally follows a power-law distribution, with an index given by $d\ln N/d\ln(R) = -z$. For the more general case where the velocity of collisions scales with size as $v_{\rm rel} \propto R^{p}$, the power-law index becomes (see eq. 8 of \citealt{cascade_velocity}):
\begin{equation} \label{eq:slope_v}
    z = \frac{3 + \xi}{6 + s - 2p},
\end{equation}
To test our code, we begin by exploring the size dependence of the encounter velocity in our fiducial simulation. The left panel of Figure \ref{fig:Qstar} shows that unless we are close to the resonance (red or purple line at 1 or 1.5 au respectively), the encounter velocity between two planetesimal of size R and 2R is dominated by $\sigma_e$ and thus remains constant for planetesimals larger than 1 km. This can be understood by noting that the forced eccentricity for $\gtrsim 1$ km-sized planetesimals shows a weak dependence on planetesimal size; see, e.g., figure 3 in Paper 1. This implies that the velocity scaling exponent is $p=0$ in this regime.

In the right panel of Figure \ref{fig:Qstar}, we can see that for sizes bigger than 1 km, the size dependence is a simple power-law, consistent with the gravity regime, where self-gravity dominates. The power-law index for this regime is $s = 1.5$. It should be noted that the power-law in the strength regime looks flatter than typically reported \citep[see][]{collisions}. This discrepancy arises because, in our case, the critical disruption energy $Q_{\rm *}$ depends on the velocity (see eq. \ref{eq:Q*}), which, itself varies with size as shown in the left panel. Finally, as illustrated in the lower panels of  Figure \ref{fig:fig1}, the inner rings initially settle to this expected slope. However, as material continues to deplete, this structure is eventually lost.

\renewcommand{\thefigure}{D\arabic{figure}}
\setcounter{figure}{0}

\section{Physical parameters tests} \label{app:physical_parameters}

\begin{figure*}\label{fig:parameters_giant}
    \includegraphics[width=1\textwidth]{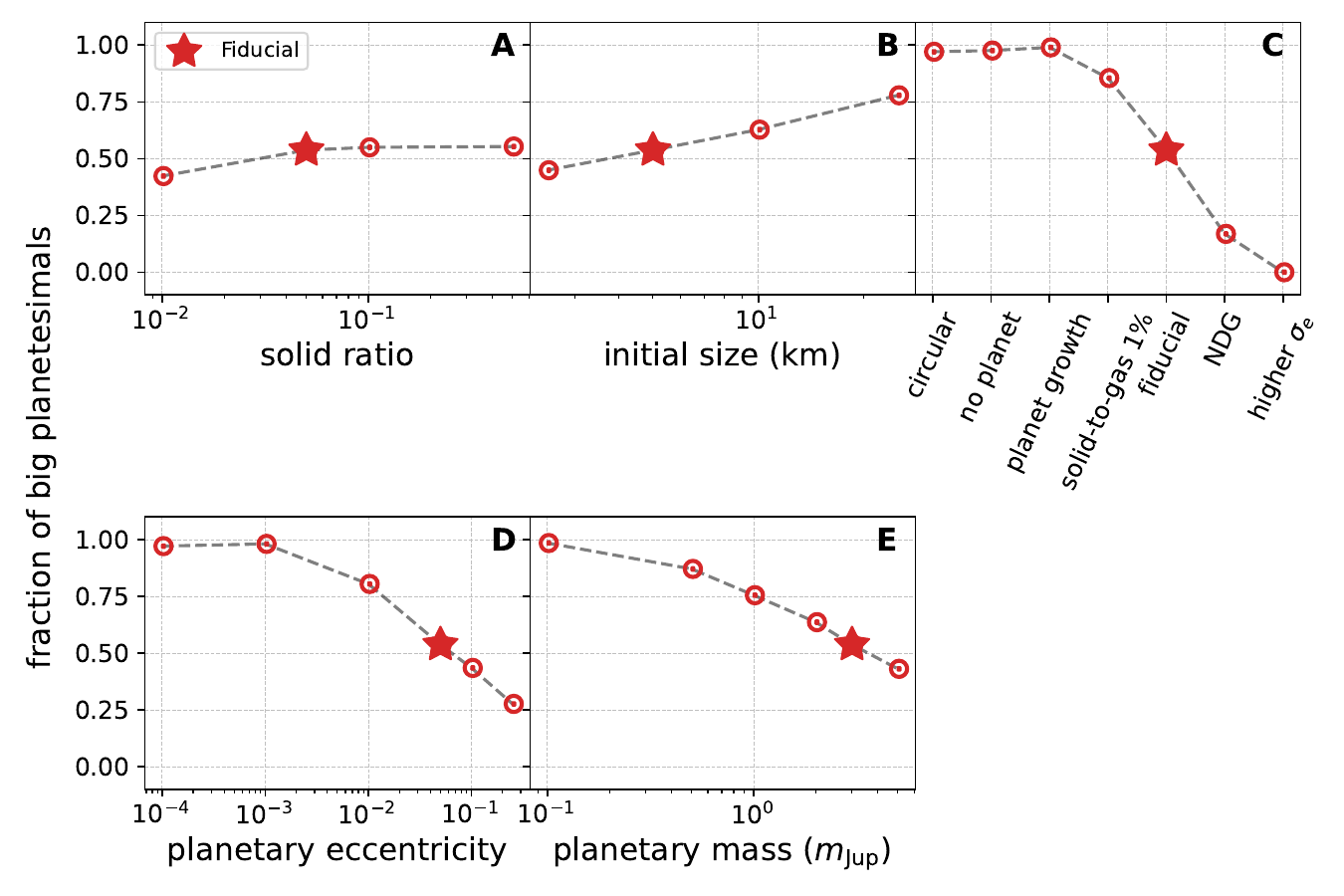}
    \caption{Each panel represents the effect of varying parameters in our tests associated with different physical assumptions. The results show the fraction of solid material that was converted into planetesimals larger than 1000 km. The fiducial case is included for reference and is marked with a star in all panels (see Table~\ref{table:models}).}
\end{figure*}

Here we present the results of a series of test where we varied different parameters--one at a time--relative to our fiducial simulation, each reflecting different physical assumptions. We classify these parameters into two categories: (1) those related to the initial distribution of solids in the disk, shown in panels A and B in Figure \ref{fig:parameters_giant}; and (2) those associated with the properties of the giant planet, shown in panels D and E in Figure \ref{fig:parameters_giant} (and discussed further in Section \ref{sec:giant_planet}). Panel C showcases a comparison between the fiducial case and some additional scenarios: the case of no disk gravity (NDG, which is further discussed in Section \ref{sec:ndg}), two different cases in which forced eccentricity of planetesimals is zero, those being a circular planet and no planet at all. This panel also includes the variations described in Section \ref{sec:other_simulations}, which explore a higher eccentricity dispersion ($\sigma_e = 2\times10^{-4}$), a case where the giant planet undergoing runaway accretion at the beginning of the simulation in a timescale of 10$^5$ years, and a case with lower solid-to-gas ratio (1\%), but higher initial gas mass (5\%$M_*$).

In panel A we examine the effect of gas-to-solids ratio. We find that above a threshold of 0.05, the fraction of material converted into large planetesimals remains constant. This suggests that increasing the initial solid content does not proportionally increase the formation of large bodies. Instead, approximately 50\% of the available material is converted, regardless of the total amount in solids. This result is important as it implies that our results can be easily scaled to disks  with different initial solids content.

Another notable trend appears in panel B of Figure \ref{fig:parameters_giant} where we observe that starting with larger planetesimals leads to a higher efficiency in the formation of 1000 km planetesimals. This outcome is expected, as larger bodies are already in the gravitational regime and are more resistant to fragmentation, reducing the amount of material lost to the fragmentation cascade. It is worth noting that despite some simulations having a large fraction of big planetesimals (panels C, D and E), this does not imply that this material is concentrated in narrow regions by the end of the simulation. See Figure \ref{fig:cdf}.

\bibliography{refs}

\end{document}